\documentclass{jpsj3}
\usepackage{txfonts}
\usepackage{bm}

\usepackage{color}

\def\gsim{\mathop {\vtop {\ialign {##\crcr 
$\hfil \displaystyle {>}\hfil $\crcr \noalign {\kern1pt \nointerlineskip } 
$\,\sim$ \crcr \noalign {\kern1pt}}}}\limits}
\def\lsim{\mathop {\vtop {\ialign {##\crcr 
$\hfil \displaystyle {<}\hfil $\crcr \noalign {\kern1pt \nointerlineskip } 
$\,\,\sim$ \crcr \noalign {\kern1pt}}}}\limits}

\title{Theory for Anomalous NMR Response in Pb$_{1-x}$Tl$_x$Te on Charge Kondo Effect}

\author{
Kazumasa Miyake$^1$\thanks{miyake@toyotariken.jp} and Hiroyasu Matsuura$^2$}
\inst{
$^1$ Center for Advanced High Magnetic Field Science, Osaka University, Toyonaka, Osaka 560-0043, 
Japan  \\
$^2$ Department of Physics, University of Tokyo, Hongo, Bunkyo-ku, Tokyo 113-0033, Japan \\

} 

\recdate
{\today}

\abst{
A theory for anomalous enhancement of NMR relaxation rate $1/T_{1}T$ of $^{125}$Te toward zero 
temperature observed in Pb$_{1-x}$Tl$_{x}$Te ($x$=0.01) is presented on the idea of the charge 
Kondo effect of valence skipping element Tl.  It is {found} that such enhancement in 
$1/T_{1}T$ is caused through enhancement of the pair-hopping 
{and inter-orbital interactions} 
between 6s electrons localized on 
Tl site and conduction electrons doped in the hole band the semiconductor PbTe, which is the 
heart of the charge Kondo effect. 
It is also found that the Knight shift $K$ is enhanced in proportion to 
{the increasing part} of 
the relaxation rate $1/T_{1}T$ in the same temperature region, implying that the Korringa relation 
does not hold in such a region of temperature. 
}

\begin{document}
\maketitle

\section{Introduction}
In the past decade, valence skipping phenomenon and related superconductivity have caused revived 
attention since the charge Kondo effect and the superconductivity had been reported in 
Pb$_{1-x}$Tl$_{x}$Te ($0.006<x<0.015$).~\cite{Matsushita}  
Since the valence state of Pb is Pb$^{2+}$, the nominal valence of Tl should be 
Tl$^{2+}$. However, the doped atom Tl is known as a valence skipping element which 
takes Tl$^{1+}$ [(6s)$^{0}$ configuration] or Tl$^{3+}$ [(6s)$^{2}$ configuration] but not Tl$^{2+}$, 
implying that (6s)$^{1}$ configuration is skipped even though a nominal valence is Tl$^{2+}$ as in 
a series of compounds.~\cite{Shannon}   The logarithmic temperature dependence in the resistivity 
$\rho(T)$ in the low temperature region ($T<10$K), which is 
robust against the magnetic field,  and the occurrence of superconductivity were shown to be
well explained on the basis of the negative-$U$ Anderson model,~\cite{Dzero} 
while fundamental properties of the negative-$U$ Anderson model had already been derived in 
the beginning of 1990s.~\cite{Taraphder}  

Recently, it has been shown by the numerical renormalization group {(NRG)} calculation that 
the pair-hopping interaction $U_{\rm ph}$ {between the} 
localized  electron and extended conduction 
electrons can give rise to the negative-$U$ or valence skipping effect and the charge Kondo effect 
simultaneously.~\cite{Matsuura}  
More explicitly, it was shown that there exist two temperature (energy) scales $T^{*}$ and 
$T_{\rm K}$ ($T_{\rm K}<T^{*}$), 
{with $T_{\rm K}$ being the Kondo temperature of the present problem,} 
i.e., at $T<T^{*}$, $(6{\rm s})^{1}$ state is excluded (skipped) 
and $(6{\rm s})^{0}$ and $(6{\rm s})^{2}$ states are degenerate 
or the negative-$U$ effect manifests itself, and at $T\lsim T_{\rm K}$, the charge Kondo effect 
occurs forming the charge singlet state, 
like the Kondo-Yosida spin singlet state.  
This origin of negative-$U$ effect was new and quite different 
from a series of proposals which had already been given,~\cite{Anderson, Yoshida, Hase, Harrison, 
Hewson, Hotta} {while its origin still remains as an active subject.~\cite{Shinzaki}  }
Since there exists no magnetic ions in Tl doped PbTe, it is reasonable 
{to expect} 
that the two-fold  charge degrees of freedom of Tl ion, Tl$^{1+}$ and Tl$^{3+}$, is the origin of 
Kondo like behavior in the resistivity so that it was called the charge Kondo 
effect.~\cite{Matsushita,Dzero,Taraphder} 

Quite recently, temperature dependence of NMR relaxation rates $1/T_{1}T$ of $^{125}$Te of 
Pb$_{1-x}$Tl$_{x}$Te were reported to exhibit diverging increase below $T=10$ K for the sample 
$x\simeq 0.01$ [ref. \citen{Mukuda}] which shows the charge Kondo effect in the resistivity and 
the superconductivity in the lower temperature region $T\lsim T_{\rm K}$.~\cite{Matsushita}  
This is non-trivial because elements consisting of this compound are all non-magnetic ones, which 
suggests that the enhancement of  $1/T_{1}T$ may give another smoking gun for the valence 
skipping or the negative-$U$ effect to play a crucial role in this compound.  

The purpose of the present paper is to clarify how the charge Kondo effect can give rise to the 
diverging behavior in $1/T_{1}T$ across the Kondo temperature $T_{\rm K}$, reinforcing that the charge 
Kondo effect is the origin of anomalous properties observed in Pb$_{1-x}$Tl$_{x}$Te ($0.006<x<0.015$).  
Organization of the paper is as follows. In Sect.\ 2,  
{a formulation for discussing the relaxation rate $1/T_{1}T$ is given on the basis of} 
the charge Kondo effect due to the pair-hopping  interaction $U_{\rm ph}$. 
In Sect.\ 3{.1},  it is shown the anomalous 
behaviors in the $1/T_{1}T$ arises from the first order process in the renormalized 
pair-hopping interaction {$U_{\rm ph}(T)$} by the charge Kondo effect at $T\gsim T_{\rm K}$. 
{
In Sect.\ 3.2, it is also shown that the $1/T_{1}T$ is similarly influenced by the renormalized inter-orbital 
interaction $U_{\rm dc}(T)$ between the localized electron and extended conduction electrons. 
As a result, it is shown in Sect.\ 3.3 that the anomalous temperature dependence of $1/T_{1}T$ 
observed in Pb$_{1-x}$Tl$_{x}$Te ($x=0.01$) is explained by these effects. 
In Sect.\ 4, it is shown that the {charge Kondo effect} also gives an enhancement of the Knight shift  $K$ 
in proportion to that of $1/T_1T$, implying that the Korringa relation is apparently broken 
where the relaxation rate is enhanced by its effect.  

\section{Formulation}
An effective model including the Coulomb interaction between the conduction electron and localized 
6s orbital (denoted by d for manifesting the relation  with the s-d model) is given by as \cite{Matsuura} 
\begin{eqnarray}
\mathcal{H}_0 = \mathcal{H}_{\rm c}+\mathcal{H}_{\rm d}  
+ \mathcal{H}_{\rm dc} +\mathcal{H}_{\rm ph} + \mathcal{H}_{\rm hyb},
\label{effmod}
\end{eqnarray}
where the first term is for the conduction electron, the second term is for 6s electrons, 
and the third and forth terms are for the Coulomb interactions $U_{\rm dc}$ and the pair-hopping 
interaction $U_{\rm ph}$ between conduction electron and localized 6s electrons. 
Explicit expression of these terms are given as
\begin{eqnarray}
\mathcal{H}_{\rm c}& \equiv&  {\frac{1}{N}}\sum_{\bf{k}\sigma}\epsilon_{\bf{k}}
c_{\bf{k}\sigma}^{\dag}c_{\bf{k}\sigma}, \label{cond} \\
\mathcal{H}_{\rm d} &\equiv& \epsilon_{d}\sum_{\sigma} n_{{\rm d}\sigma}, \label{dlevel}\\
\mathcal{H}_{\rm dc} &\equiv&  U_{\rm dc}\,{\frac{1}{N}}\sum_{\bf{k}  \sigma} 
c_{\bf{k}\sigma}^{\dagger}c_{\bf{k}\sigma}n_{{\rm d}\sigma^{\prime}},  \\
\mathcal{H}_{\rm ph} &\equiv&  U_{\rm ph}\,{\frac{1}{N}}\sum_{\bf{k} \bf{k}^{\prime}}
\left(d_{\uparrow}^{\dagger}d_{\downarrow}^{\dagger} c_{\bf{k}\downarrow} 
c_{\bf{k^{\prime}\uparrow}}+{\rm h.c.}\right),
\\
\mathcal{H}_{\rm hyb} &\equiv& V_{\rm dc}\,{\frac{1}{{\sqrt{N}}}}\sum_{\bf{k}\sigma} (c_{\bf{k}\sigma}^{\dagger}d_{\sigma}+{\rm h.c.}), 
\label{hybrid} 
\end{eqnarray}
where  {$N$ is the number of lattice sites, and} $n_{{\rm d}\sigma}\equiv d^{\dagger}_{\sigma}d_{\sigma}$ is the number operator of the localized 
6s electrons.  
Hereafter, the origin of energy is taken as the Fermi energy of conduction electrons 
$\epsilon_{\rm F}$, the chemical potential at $T=0$, and the temperature $T$ is assumed to be low 
enough compared to $\epsilon_{\rm F}$, i.e., ${k_{\rm  B}}T\ll \epsilon_{\rm F}$. 

As discussed in Ref.\ \citen{Matsuura}, the pair-hopping interaction $U_{\rm ph}$ can stabilize 
the valence skipping state and cause the charge Kondo effect under certain condition. 
The origin of this phenomenon 
can be understood intuitively if we note that the $U_{ph}$ is transformed to the pseudo-spin 
flipping exchange interaction (the origin of the Kondo effect) 
by the particle-hole transformation for the annihilation operators $d_{\downarrow}$ and 
$c_{{\bf k}\downarrow}$ as shown explicitly in Appendix\ref{equivalence}. 

The NMR relaxation rate $1/T_{1}$ is given by the Moriya formula as \cite{Moriya}
\begin{equation}
\frac{1}{T_{1}T}=A^{2}\frac{1}{\omega}{\rm Im}
\Gamma^{\rm R}(\omega+{\rm i}\delta),
\label{rates0}
\end{equation}
where $A$ is the hyper-fine coupling constant between electron and nuclei, 
{and 
$\Gamma({\rm i}\omega_{\nu})$ is the transverse spin susceptibility of conduction electrons 
at certain Te site where NMR relaxation is observed and has several contributions, in general.  
The $\Gamma_{\rm ph}({\rm i}\omega)$ arising from the lowest order process in $U_{\rm ph}$ 
is given by the Feynman diagram shown in Fig.\ \ref{Fig:1} and its vertical inversion as follows:}
\begin{eqnarray}
& & 
\Gamma_{\rm ph}({\rm i}\omega_{\nu})
= 2V_{\rm dc}^{2}T^{{2}}\sum_{\varepsilon_{n}}
U_{\rm ph}
G_{\rm c}({\bf r}_{ij},{\rm i}\varepsilon_{n})
G_{\rm c}({\bf r}_{ij},{\rm i}\varepsilon_{n}+{\rm i}\omega_{\nu})G_{\rm c}({\bf r}_{ij},-{\rm i}\varepsilon_{n})
G_{\rm c}({\bf r}_{ij},-{\rm i}\varepsilon_{n}+{\rm i}\omega_{\nu})
\nonumber
\\
& &\qquad\qquad\qquad\qquad\qquad
\times
G_{\rm d}(-{\rm i}\varepsilon_{n}+{\rm i}\omega_{\nu})G_{\rm d}({\rm i}\varepsilon_{n}),
\label{rate1}
\end{eqnarray}
where we have used the property 
$G_{\rm c}(-{\bf r}_{ij},{\rm i}\varepsilon_{n})=G_{\rm c}({\bf r}_{ij},{\rm i}\varepsilon_{n})$ etc.,
{
and the factor 2 arises from the diagrams of the vertical inversion.  
}

The expression [Eq.\ \ref{rate1}] is verified by the Wick decomposition as 
\begin{eqnarray}
& &
\left\langle 
T_{\tau} \left[{\bar c}_{i\uparrow}(\tau)c_{i\downarrow}(\tau)(-U_{\rm ph})
{\bar d}_{{j}\uparrow}(\tau^{\prime})
{\bar d}_{j\downarrow}(\tau^{\prime})
{c}_{j\downarrow}(\tau^{\prime})c_{j\uparrow}(\tau^{\prime})
{\bar c}_{i\downarrow}(\tau^{\prime\prime})c_{i\uparrow}(\tau^{\prime\prime})
\right]
\right\rangle
\nonumber
\\
& &\quad
=U_{\rm ph}\langle T_{\tau}c_{j\uparrow}(\tau^{\prime}){\bar c}_{i\uparrow}(\tau)\rangle
\langle T_{\tau}c_{j\downarrow}(\tau^{\prime}){\bar c}_{i\downarrow}(\tau^{\prime\prime})\rangle
\langle T_{\tau}c_{i\downarrow}(\tau){\bar d}_{j\downarrow}(\tau^{\prime})\rangle
\langle 
T_{\tau}c_{i\uparrow}(\tau^{\prime\prime}){\bar d}_{j\uparrow}(\tau^{\prime})
\rangle
\label{Wick1}
\end{eqnarray}


The reason why the Green functions with $\mp{\rm i}\varepsilon_{n}$ and 
$\pm{\rm i}\varepsilon_{n}+{\rm i}\omega_{\nu}$ are paired in Fig.\ \ref{Fig:1} is 
based on the fact 
that the Kondo-like renormalization enhancing the pair-hopping interaction $U_{\rm ph}$ arises 
for the 
{
annihilation process of pair of conduction electrons with ${\rm i}\varepsilon_{n}$ and 
$-{\rm i}\varepsilon_{n}$ as} 
discussed in Appendix\ref{equivalence}. Namely, the expression [Eq.\ (\ref{rate1})] is regarded as 
the most 
{
divergent part when $U_{\rm ph}$ divergently grows for the elestic scattering channel by 
the charge Kondo effect as decreasing temperature.  
This treatment of extracting the most divergent contribution is analogous to 
that adopted in the problem of estimating the effect of 
superconducting fluctuations to the conductivity near the superconducting 
transition point.~\cite{Maki,Thompson, AL}
}
The renormalization of $U_{\rm ph}$ for a specified localized electron 
arises through the higher order terms in $U_{\rm ph}$ as 
in the conventional Kondo effect, and can be performed by the renormalization group {(RG)} method such as 
the poorman's scaling approach as discussed below.~\cite{PoormanScaling} 
On the other hand, 
the higher order terms in $U_{\rm ph}$ among different localized electrons are higher order 
in the impurity concentration and are  
safely neglected in the present case where the concentration of Tl impurity is small about 10$^{-2}$.  

\begin{figure}[h]
\begin{center}
\rotatebox{0}{\includegraphics[angle=0,width=0.6\linewidth]{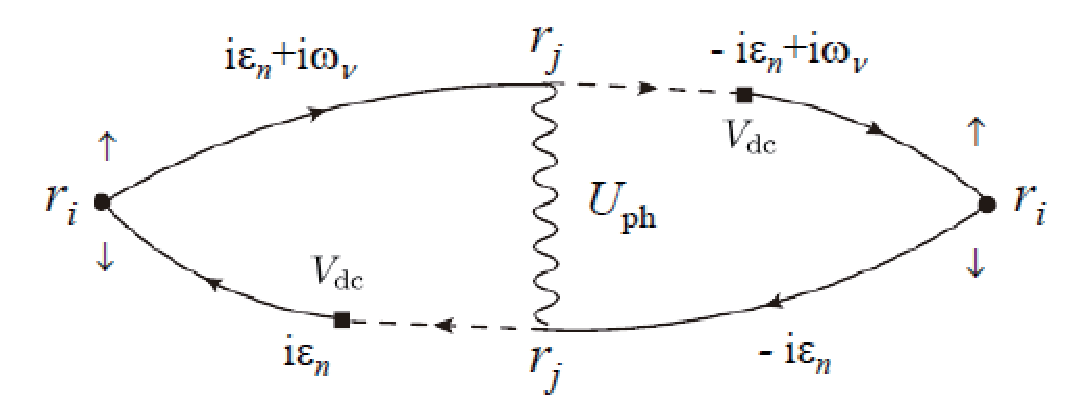}}
\caption{
Feynman diagram giving the NMR longitudinal relaxation rates $1/T_{1}T$ at Te (${\bf r}_{i}$) site 
in the lowest order with respect to the pair-hopping interaction $U_{\rm ph}$. 
Wavy line represents $U_{\rm ph}$ at Tl (${\bf r}_{j}$) site. 
Solid lines with arrow and 
dashed lines with arrow represent the Matsubara Green function of conduction electrons of conduction 
band (in hole picture) and 6s electron at Tl site, respectively. Filled squares represent the hybridization 
$V_{\rm dc}$ between conduction electrons and 6s electron at Tl site. 
}
\label{Fig:1}
\end{center}
\end{figure}

As shown in Appendix\ref{SpectralFunction}, the Green function of the conduction electrons 
$G_{\rm c}({\bf r},{\rm i}\varepsilon_{n})$ is 
expressed by a spectral representation as 
\begin{equation}
G_{\rm c}({\bf r},{\rm i}\varepsilon_{n})=\int_{-\infty}^{\infty}{\rm d}y\,
\frac{\rho({\bf r},y)}{{\rm i}\varepsilon_{n}-y}, 
\label{GF1}
\end{equation} 
with a spectral function 
\begin{equation}
\rho({\bf r},y)
=N_{\rm F}\frac{e^{-(r/2\ell)}}{k_{\rm F}r}
\sin\left[\sqrt{\frac{y}{\epsilon_{\rm F}}+1}\,(k_{\rm F}r)\right]
\theta(y+\epsilon_{\rm F})\,\theta(\epsilon_{\rm c}-y),
\label{GF2}
\end{equation} 
where $N_{\rm F}\equiv mk_{\rm F}/2\pi^{2} {N}$ is the density of states of conduction electrons 
at the Fermi level,  {per lattice site and spin component, and} 
$\epsilon_{\rm F}$ and $\epsilon_{\rm c}$ are the Fermi energy of conduction electrons and 
the energy cutoff of the conduction band in the hole (check) picture and a mean-free path due to 
the impurity scattering,~\cite{AGD} respectively.   
In the limit $k_{\rm F}r\ll1$, the spectral function takes a form as 
\begin{equation}
\rho({\bf r},y)\approx
 {N_{\rm F}}\,e^{-(r/2\ell)}\sqrt{\frac{y}{\epsilon_{\rm F}}+1}\,
\theta(y+\epsilon_{\rm F})\,\theta(\epsilon_{\rm c}-y). 
\label{GF3}
\end{equation} 
The Green function of localized electron at valence-skipping site, i.e., Tl site, is given by 
\begin{equation}
G_{\rm d}({\rm i}\varepsilon_{n})=\frac{1}{{\rm i}\varepsilon_{n}-\epsilon_{\rm d}}
\label{GF4}
\end{equation}
where $\epsilon_{\rm d}$ is the energy level of localized electron measured from the chemical potential.

\section{NMR Relaxation Rate at $T\gsim T_{\rm K}$}
{
\subsection{Effect of pair-hopping interaction}
In this subsection, the NMR relaxation rate triggered by the pair-hopping interaction $U_{\rm ph}$.  
}  
As shown in Appendix\ref{Gamma},  
${\rm Im}\Gamma^{\rm R}(\omega+{\rm i}\delta)$ is given by 
\begin{eqnarray}
& &
{\rm Im}\Gamma_{\rm ph}^{\rm R}(\omega+{\rm i}\delta)
=-\frac{ {2}\pi V_{\rm dc}^{2}{T}U_{\rm ph}}{\epsilon_{\rm d}^{2}}
\int_{-\infty}^{\infty}{\rm d}y_{4}\,
\left[{\rm th}\left(\frac{y_{4}-\omega}{2T}\right)-{\rm th}\frac{y_{4}}{2T}\right]
\nonumber
\\
& &
\qquad\qquad\qquad\qquad\quad
\times
\left[\rho({\bf r}_{ij},y_{4}+\omega)\rho({\bf r}_{ij},-y_{4}+\omega)
G_{\rm c}^{\prime{\rm R}}({\bf r}_{ij},-y_{4})G_{\rm c}^{\prime{\rm R}}({\bf r}_{ij},y_{4})
\right]. 
\label{Gamma1}
\end{eqnarray}
Therefore, to the leading order in $\omega$ and  in the low temperature 
limit, $T\ll \epsilon_{\rm F}$, ${\rm Im}\Gamma^{\rm R}(\omega+{\rm i}\delta)/\omega$ 
is expressed in a compact form as
\begin{equation}
\frac{\displaystyle {\rm Im}\Gamma_{\rm ph}^{\rm R}(\omega+{\rm i}\delta)}{\omega}\approx
\frac{{4}\pi V_{\rm dc}^{2}{T}U_{\rm ph}}{\epsilon_{\rm d}^{2}}
\left[\rho({\bf r}_{ij},0)G_{\rm c}^{\prime{\rm R}}({\bf r}_{ij},0)\right]^{2}.
\label{rates7D}
\end{equation}
With the use of definition Eq.\ (\ref{GF2}) for the spectral function $\rho({\bf r},y)$, 
$\rho({\bf r},0)$ is given by 
\begin{equation}
\rho({\bf r},0)
=N_{\rm F}\frac{e^{-(r/2\ell)}}{k_{\rm F}r}
\sin\,(k_{\rm F}r),
\label{GF2A}
\end{equation} 
and an explicit form of 
$G_{\rm c}^{\prime{\rm R}}({\bf r},\varepsilon)$ is given by 
\begin{equation} 
G_{\rm c}^{\prime{\rm R}}({\bf r},\varepsilon)
=N_{\rm F}\frac{e^{-(r/2\ell)}}{k_{\rm F}r}
\int_{-\epsilon_{\rm F}}^{\epsilon_{\rm c}}{\rm d}y\sin\left[\sqrt{\frac{y}{\epsilon_{\rm F}}+1}\,(k_{\rm F}r)\right]\,
\frac{1}{\varepsilon-y}.
\label{GF1B}
\end{equation}
Therefore, $G_{\rm c}^{\prime{\rm R}}({\bf r},0)$ is expressed as 
\begin{equation} 
G_{\rm c}^{\prime{\rm R}}({\bf r},0)=-N_{\rm F}\,e^{-(r/2\ell)}J(k_{\rm F}r),
\label{GF1C}
\end{equation}
with a function defined as 
\begin{equation} 
J(k_{\rm F}r) \equiv
\frac{1}{k_{\rm F}r}
\int_{-\epsilon_{\rm F}}^{\epsilon_{\rm c}}{\rm d}y\sin\left[\sqrt{\frac{y}{\epsilon_{\rm F}}+1}\,(k_{\rm F}r)\right]\,
\frac{1}{y}.
\label{GF1D}
\end{equation} 
The result of numerical integration in Eq.\ (\ref{GF1D}) is shown in Fig.\ \ref{Fig:JkFr} for a series of 
$(\epsilon_{\rm c}/\epsilon_{\rm F})$s.  

\begin{figure}[h]
\begin{center}
\rotatebox{0}{\includegraphics[angle=0,width=0.7\linewidth]{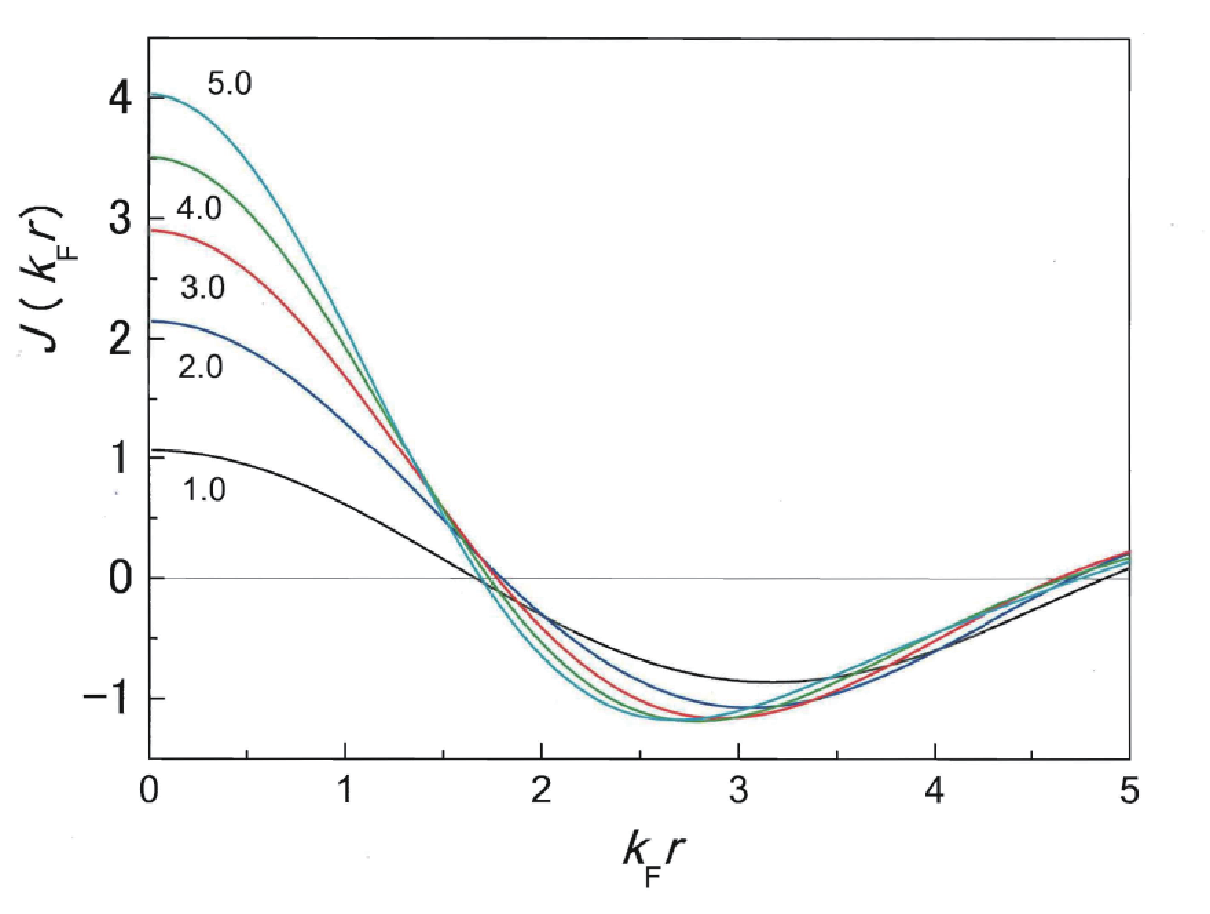}}
\caption{
$J(k_{\rm F}r)$ for a series of $(\epsilon_{\rm c}/\epsilon_{\rm F})$s.
 }
\label{Fig:JkFr}
\end{center}
\end{figure}

Substituting Eqs.\ (\ref{GF2A}) and (\ref{GF1B}) into Eq.\ (\ref{rates7D}), the NMR relaxation rate 
$(1/T_{1}T)_{\rm ph}$ [Eq.\ (\ref{rates0})] in the low temperature limit $(T\ll \epsilon_{\rm F})$ is given by 
\begin{equation}
\left(\frac{1}{T_{1}T}\right)_{\rm ph}\approx A^{2}
\frac{4\pi  {N_{\rm F}^{2}}(V_{\rm dc}N_{\rm F})^{2}TU_{\rm ph}}
{\epsilon_{\rm d}^{2}}\,e^{-(r/\ell)}
\left[
\frac{\sin(k_{\rm F}r)}{k_{\rm F}r}\,J(k_{\rm F}r)
\right]^{2}.
\label{rates7E}
\end{equation}
This formula offers the basis for discussing 
anomalous enhancement of the relaxation rate $1/T_{1}T$ in the region $T{\gsim} T_{\rm K}$. 

In the limit $k_{\rm F}r\ll 1$, the integration with respect to $y$ in Eq.\ (\ref{GF1D}) 
is given by 
\begin{equation}  
J(k_{\rm F}r) \simeq 2\sqrt{\displaystyle \frac{\epsilon_{\rm F}}{\epsilon_{\rm c}}+1}
+
\log\left|\frac{\sqrt{\epsilon_{\rm c}+\epsilon_{\rm F}}-\sqrt{\epsilon_{\rm F}}}
{\sqrt{\epsilon_{\rm c}+\epsilon_{\rm F}}+\sqrt{\epsilon_{\rm F}}}
\right|,
\label{GF1E}
\end{equation} 
as shown in Appendix\ref{RealPart}.  
On the other hand, in the limit $k_{\rm F}r\gg 1$, the asymptotic form of $J(k_{\rm F}r)$ 
is given as 
\begin{equation} 
J(k_{\rm F}r) \approx
\frac{1}{k_{\rm F}r}\pi\cos\,(k_{\rm F}r),
\label{GF1F}
\end{equation} 
as shown in Appendix\ref{JkFr}. Note that the asymptotic form shown in Fig.\ \ref{Fig:JkFr} 
is consistent with the result [Eq.\ (\ref{GF1F})].  
Therefore, in the limit $k_{\rm F}r\gg 1$, $1/T_{1}T$ 
given by Eq.\ (\ref{rates7E}) is in proportion to 
$e^{-(r/\ell)}[\sin\,(2k_{\rm F}r)/(k_{\rm F}r)^{2}]^{2}$. 


According to the result based on the NRG calculation,~\cite{Matsuura} the renormalized 
pair-hopping interaction $U_{\rm ph}$, owing to the impurity charge Kondo effect, 
is expected to exhibit a diverging $T$ dependence as $T$ decreases.  
This is because, as shown in Appendix\ref{equivalence}, the $U_{\rm ph}$ is transformed to 
the spin exchange interaction by the particle-hole transformation for the down spin component 
of both localized (d) and conduction electrons, so that it is enhanced in parallel to the magnetic 
Kondo effect. 
Indeed, the renormalization of $U_{\rm ph}$ up to the second order in $U_{\rm ph}$ and 
$U_{\rm dc}$ is given by the Feynman diagrams shown in Fig.\ \ref{2ndOrderProcess}(a). 
Similarly, that of $U_{\rm dc}$ is given by the Feynman diagram shown in 
Fig.\ \ref{2ndOrderProcess}(b). These processes are formally the same as those appearing the magnetic Kondo problem because $U_{\rm ph}$ and $U_{\rm dc}$ correspond to $J_{\perp}/2$ and $J_{z}/4$ in the 
anisotropic s-d model, respectively, in the mapped world by the transformations 
[Eqs.\ (\ref{Shiba1}) and (\ref{Shiba2})] as discussed in Appendix\ref{equivalence}.

\begin{figure}[h]
\begin{center}
\rotatebox{0}{\includegraphics[angle=0,width=0.8\linewidth]{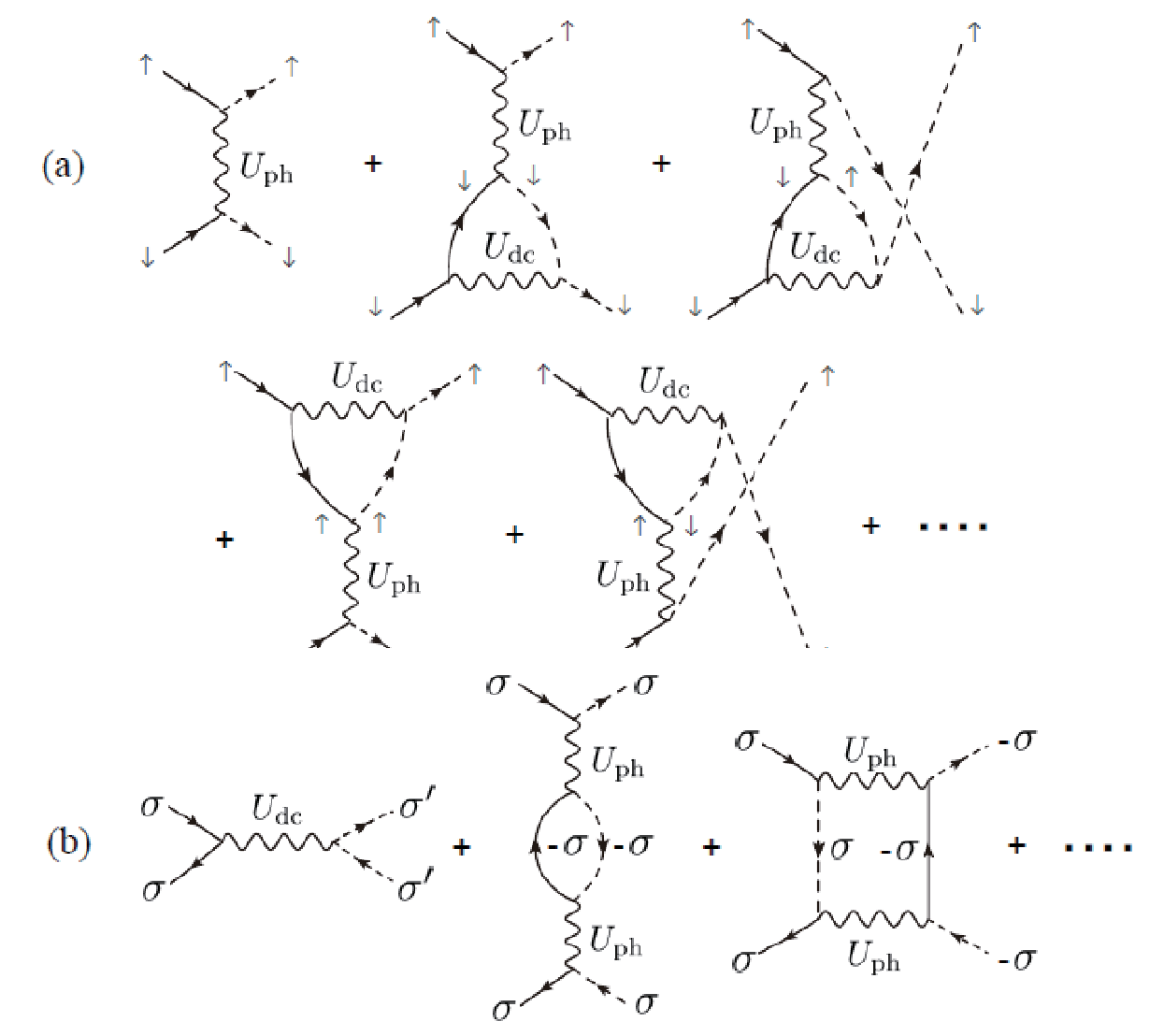}}
\caption{
Feynman diagram for the renormalization of (a) the pair-hopping interaction $U_{\rm ph}$ 
and (b) the inter-orbital interaction $U_{\rm dc}$, up to the second order in in $U_{\rm ph}$ and 
$U_{\rm dc}$. These are formally the same as those in the anisotropic s-d model. 
}
\label{2ndOrderProcess}
\end{center}
\end{figure}

In order to take into account a series of higher order corrections with respect to $U_{\rm ph}$ and 
$U_{\rm dc}$, it is useful to rely on RG approaches in general. 
{
For example, this type of approach has been successfully applied to understand the anomalous 
temperature dependence of the valence of the Sm ion in a filled-Skutterudite compound 
SmOs$_4$Sb$_{12}$~\cite{Tanikawa}.
}
Here, we adopt the one-loop order  (or poorman' scaling) approximation.~\cite{PoormanScaling} 
As shown in Apendix\ref{Scaling} [Eq.\ (\ref{E8})], 
the $T$ dependent renormalized pair-hopping interaction 
$U_{\rm ph}(T)\,{[\equiv y_{\perp}(T)/2N_{\rm F}]}$  
{
is given as
\begin{equation} 
U_{\rm ph}(T)=\frac{1}{2N_{\rm F}\log(T/T_{\rm K})},
\label{GF1H}
\end{equation}
and
}
has the logarithmic $T$ dependenc{e, in the high temperature region} 
at $T\gsim T_{\rm K}$, like 
\begin{equation}
U_{\rm ph}(T)\approx U_{\rm ph}^{0}
-4N_{\rm F}U_{\rm ph}^{0}U_{\rm dc}^{0}\,\log\frac{T}{E_{\rm c}^{0}},
\label{rates7F}
\end{equation}
where $U_{\rm ph}^{0}$ and $U_{\rm dc}^{0}$ are the bare pair-hopping and inter-orbital interactions, 
respectively, and $E_{\rm c}^{0}$ is the bare bandwidth of conduction electrons. 
{
On the other hand, $U_{\rm ph}(T)$ exhibits divergent behavior toward $T=T_{\rm K}$ as 
$U_{\rm ph}(T)\approx U_{\rm ph}^{0}/\log(T/T_{\rm K)}$ in the one-loop order 
RG approximation.  }
Replacing $U_{\rm ph}$ in Eq.\ (\ref{rates7E}) by $U_{\rm ph}(T)$ [Eq.\ (\ref{GF1H})], 
the NMR relaxation rates $(1/T_{1}T)_{\rm ph}$ at Te site (${\bf r}$) is given by
\begin{equation}
\left(\frac{1}{T_{1}T}\right)_{\rm ph} \approx A^{2}
\frac{4\pi  {N_{\rm F}^{2}}(V_{\rm dc}N_{\rm F})^{2}}{\epsilon_{\rm d}^{2}}\,e^{-(r/\ell)}
\left[
\frac{\sin(k_{\rm F}r)}{k_{\rm F}r}\,J(k_{\rm F}r)
\right]^{2}
{TU_{\rm ph}(T)}.
\label{rates7G}
\end{equation}

The procedure of replacing the bare pair-hopping interaction $U_{\rm ph}^{0}$ by the renormalized one 
$U_{\rm ph}(T)$ may be justified by the expression [Eq. \ (\ref{Gamma1})] in which major 
contribution comes from the conduction electrons with the energy $y_{4}\lsim T$. 
Equation (\ref{rates7G}) is one of central results of the present paper. 
Namely, 
the NMR relaxation rates $1/T_{1}T$ at Te sites near the Tl site should 
exhibit pronounced increase as $T$ decreases toward the Kondo temperature $T_{\rm K}$ 
of the charge Kondo  effect.  
This result { is a signature of} the diverging increase in the NMR relaxation rate 
$1/T_{1}T$ of $^{125}$Te in Pb$_{1-x}$Tl$_{x}$Te observed below $T=10$ K for the sample 
$x\simeq 0.01$ in Ref.\ \citen{Mukuda}.  
Note that $1/T_{1}T$ of $^{125}$Te in Pb$_{1-x}$Na$_{x}$Te with non-valence skipping 
element Na does not exhibit such enhancement,~\cite{Horikawa} suggesting that the valence skipping effect is 
the origin of such enhancement.   
The result is expected to remain essentially valid if we adopt more solid calculations, 
such as the NRG calculation,~\cite{Matsuura} 
because the diverging behavior in the renormalized pair-hopping interaction $U_{\rm ph}$ 
toward $T=T_{\rm K}$ is still expected {as discussed in the end of the present section.}  

Concluding this subsection, it should be remarked that there exist higher order corrections in $U_{\rm ph}$ 
to the diagram shown in Fig.\ \ref{Fig:1} which is essentially independent of the Kondo-like 
renormalization on the pair-hopping interaction $U_{\rm ph}$ itself given by the vertical processes 
shown in Fig.\ \ref{2ndOrderProcess}(a).   
For example, such a next order correction $\Delta U_{\rm ph}({\rm i}\omega_{\nu})$ to $U_{\rm ph}$ 
in Fig.\ \ref{Fig:1} (in the horizontal direction) is given by 
Fig.\ \ref{Fig:2} whose analytic expression is  
\begin{eqnarray}
& &
\Delta U_{\rm ph} ({\rm i}\omega_{\nu})=-U_{\rm ph}^{2}T\sum_{\varepsilon_{n^{\prime\prime}}}
G_{\rm d}({\rm i}\varepsilon_{n^{\prime\prime}}+{\rm i}\omega_{\nu})
G_{\rm d}(-{\rm i}\varepsilon_{n^{\prime\prime}})
\nonumber
\\
& &
\qquad\qquad\,\,\,
=-U_{\rm ph}^{2}\frac{1}{2\epsilon_{\rm d}-{\rm i}\omega_{\nu}}
{\rm tanh}\left(\frac{\epsilon_{\rm d}}{2T}\right), 
\label{VertexCorrection1}
\end{eqnarray}
{
where the minus sign arises from the order of perturbation expansion with respect to $U_{\rm ph}$ 
compared to the first order term in $U_{\rm ph}$ given by Fig.\ \ref{Fig:1}. 
}  
After analytic continuation ${\rm i}\omega_{\nu}\to \omega+{\rm i}\delta$, 
$\Delta U_{\rm ph}^{\rm R} (\omega+{\rm i}\delta)$ is reduced to  
\begin{equation}
\Delta U_{\rm ph}^{\rm R} (\omega+{\rm i}\delta)
=-U_{\rm ph}^{2}\frac{1}{2\epsilon_{\rm d}}
{\rm tanh}\left(\frac{\epsilon_{\rm d}}{2T}\right),
\label{VertexCorrection2}
\end{equation}
where we have used the relation $\delta(\omega-\epsilon_{\rm d})=0$ which holds at $\omega\sim 0$. 
This correction is negative and gives the suppression of the effect of the pair-hopping interaction 
in contrast to the enhancement by the Kondo-like renormalization given by the vertical processes 
shown in 
Fig.\ \ref{2ndOrderProcess}. This kind of counter renormalization effect is a general aspect 
of the Kondo effect in which the effect of the divergent increase of the effective exchange 
coupling constant $J$ finally becomes inactive because of the Kondo-Yosida singlet 
formation~\cite{Yosida,Yoshimori} by the divergent exchange coupling constant 
itself.~\cite{PoormanScaling} 
Indeed, it was demonstrated that the vertex correction for the spin susceptibility is crucial to 
obtain the Korringa relation characteristic of the local Fermi liquid property in multi 
orbital d-electron impurity Anderson model.~\cite{ShibaKorringa} 
However, such an effect of renormalization becomes crucial only at $T<T_{\rm K}$ where 
the Kondo-Yosida singlet state is formed.  Therefore, it plays minor roles in the region 
$T\gsim T_{\rm K}$ where the diverging $T$ dependence in $1/T_{1}T$ is observed.

\begin{figure}[h]
\begin{center}
\rotatebox{0}{\includegraphics[angle=0,width=0.6\linewidth]{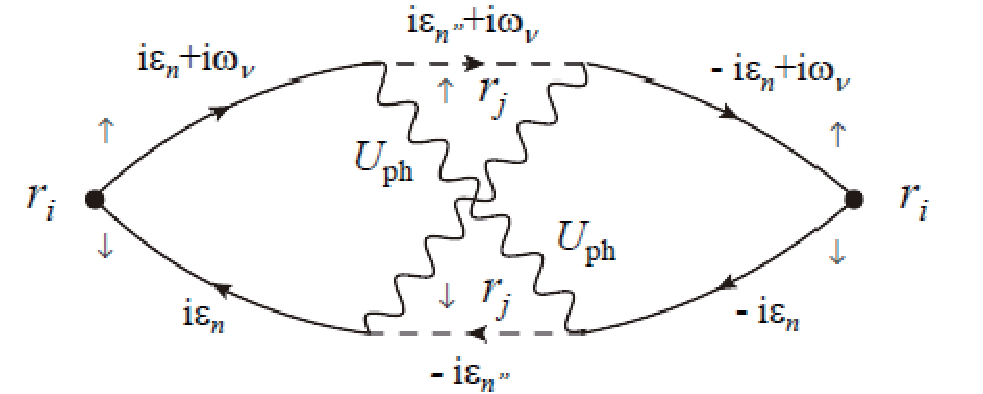}}
\caption{
Feynman diagram giving the NMR longitudinal relaxation rates $1/T_{1}T$ at Te (${\bf r}_{i}$) site 
in the second order in the pair-hopping interaction $U_{\rm ph}$. 
Notations are the same as those of Fig.\ \ref{Fig:1} 
}
\label{Fig:2}
\end{center}
\end{figure}

{
\subsection{Effect of inter-orbital interaction}
In this subsection, the NMR relaxation rate triggered by the inter-orbital interaction $U_{\rm dc}$.   
Although it was demonstrated that the pair-hopping interaction $U_{\rm ph}$ enhances the NMR 
relaxation rate toward $T=T_{\rm K}$, 
}
{
it is crucial to note that the inter-orbital interaction $U_{\rm dc}$ is also renormalized 
(enhanced) by the charge Kondo effect, as shown in Appendix\ref{Scaling} [Eq.\ (\ref{E9})], and 
the $T$ dependent $U_{\rm dc}(T)$ also has the logarithmic $T$ 
dependence in the high temperature region $T\gsim T_{\rm K}$ as 
\begin{equation}
U_{\rm dc}(T)
\approx U_{\rm dc}^{0}-N_{\rm F}(U_{\rm ph}^{0})^{2}\,\log\frac{T}{E_{\rm c}^{0}},
\label{rates7H}
\end{equation}
}
{ and exhibits divergent behavior toward $T=T_{\rm K}$ as 
$U_{\rm dc}(T)\approx 1/[4N_{\rm F}\log(T/T_{\rm K})]$ in the one-loop order RG approximation, 
like in Eq.\ (\ref{GF1H}) as shown in Appendix \ref{Scaling}.
}
{
Therefore, we have to keep the relaxation processes caused by the effect of $U_{\rm dc}$.  
There are three types of processes causing the relaxation in the first order in $U_{\rm dc}$. 
One of them is 
given by  the Feynman diagram shown in Fig.\ \ref{Fig:Udc_V} or its vertical inversion. 
This is a type of vertex correction to the local magnetic susceptibility of conduction electrons 
at certain Te site. Corresponding to the expression 
[Eq.\ (\ref{rate1})], the analytic expression for the function $\Gamma_{\rm dcV}({\rm i}\omega_{\nu})$ 
for this correction is given by 
\begin{equation}
\Gamma_{\rm dcV}({\rm i}\omega_{\nu})
= \frac{2V_{\rm dc}^{2}}{\epsilon_{\rm d}^{2}}T^{{2}}\sum_{\varepsilon_{n}}
U_{\rm dc}
\left[G_{\rm c}({\bf r}_{ij},{\rm i}\varepsilon_{n})
G_{\rm c}({\bf r}_{ij},{\rm i}\varepsilon_{n}+{\rm i}\omega_{\nu})\right]^{2},
\label{rate1V}
\end{equation}
where we have used the property 
$G_{\rm c}(-{\bf r}_{ij},{\rm i}\varepsilon_{n})=G_{\rm c}({\bf r}_{ij},{\rm i}\varepsilon_{n})$ etc., 
and the factor 2 arises from the diagram of the {inversion of upside down.} 

\begin{figure}[h]
\begin{center}
\rotatebox{0}{\includegraphics[angle=0,width=0.6\linewidth]{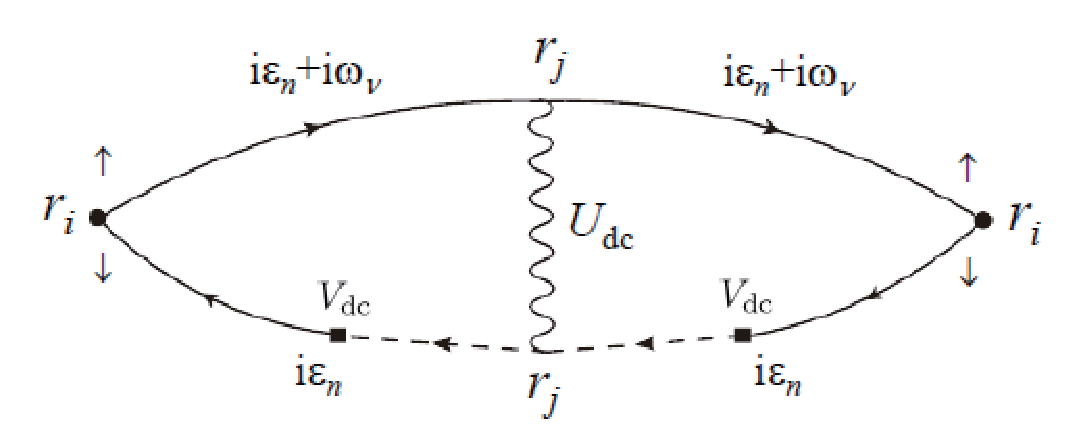}}
\caption{
Feynman diagram giving the NMR longitudinal relaxation rates $1/T_{1}T$ at Te (${\bf r}_{i}$) site 
in the lowest order with respect to the inter-orbital interaction $U_{\rm dc}$ which corresponds to 
that given by Fig.\ \ref{Fig:1}.  
{The other diagram is given by the inversion of upside down.} 
}
\label{Fig:Udc_V}
\end{center}
\end{figure}

Other types of processes causing the relaxation are given by the Feynman diagrams shown in 
Figs.\  \ref{Fig:Udc_F}(a) and (b) and Figs.\  \ref{Fig:Udc_H}(a) and (b).  These are a type of 
the self-energy corrections to the conduction electrons. 
{
It is easy to see that the contribution from Figs.\  \ref{Fig:Udc_H}(a) and (b) {is} twice of 
that from Figs.\  \ref{Fig:Udc_F}(a) and (b), due to spin degrees of freedom, 
with negative sign due to the extra Fermion-loop factor $(-1)$.   
}
{
Therefore, the analytic expression for the function $\Gamma_{\rm dcS}({\rm i}\omega_{\nu})$ 
corresponding to diagrams shown in Figs.\  \ref{Fig:Udc_F}(a) and (b) and Figs.\  \ref{Fig:Udc_H}(a) and (b) 
is given as 
\begin{eqnarray}
& & 
\Gamma_{\rm dcS}({\rm i}\omega_{\nu})
=-\frac{2V_{\rm dc}^{2}}{\epsilon_{\rm d}^{2}}T^{{2}}\sum_{\varepsilon_{n}}
U_{\rm dc}\Biggl\{
[G_{\rm c}({\bf r}_{ij},{\rm i}\varepsilon_{n})]^{2}
G_{\rm c}(0,{\rm i}\varepsilon_{n})G_{\rm c}(0,{\rm i}\varepsilon_{n}-{\rm i}\omega_{\nu})
[G_{\rm d}({\rm i}\varepsilon_{n})]^{2}
\nonumber
\\
& &
\qquad\qquad\qquad\qquad\qquad\,\,
+[G_{\rm c}({\bf r}_{ij},{\rm i}\varepsilon_{n})]^{2}
G_{\rm c}(0,{\rm i}\varepsilon_{n})G_{\rm c}(0,{\rm i}\varepsilon_{n}+{\rm i}\omega_{\nu})
[G_{\rm d}({\rm i}\varepsilon_{n})]^{2}
\Biggr\},
\label{rates7I}
\end{eqnarray} 
where the first and second terms are for the Figs.\ \ref{Fig:Udc_F}(a) and (b), respectively, and 
the factor 2 arises from the diagram {of the inversion of upside down.}  
}

\begin{figure}[h]
\begin{center}
\rotatebox{0}{\includegraphics[angle=0,width=0.6\linewidth]{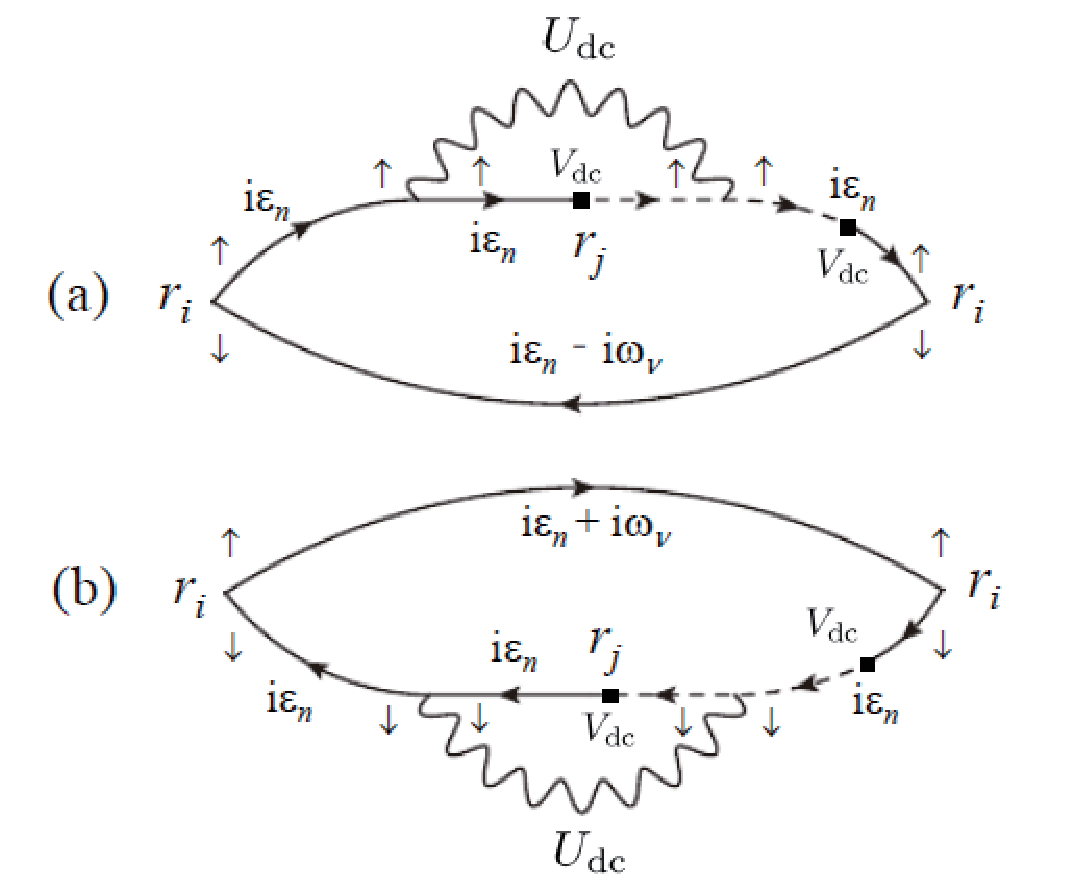}}
\caption{
Feynman diagram giving the NMR longitudinal relaxation rates $1/T_{1}T$ at Te (${\bf r}_{i}$) site 
in the lowest order with respect to the inter-orbital interaction $U_{\rm dc}$ which corresponds to 
a Fock type self-energy correction.  
{The other diagrams are given by the mirror inversion with respect to $U_{\rm dc}$.} 
}
\label{Fig:Udc_F}
\end{center}
\end{figure}

\begin{figure}[h]
\begin{center}
\rotatebox{0}{\includegraphics[angle=0,width=0.6\linewidth]{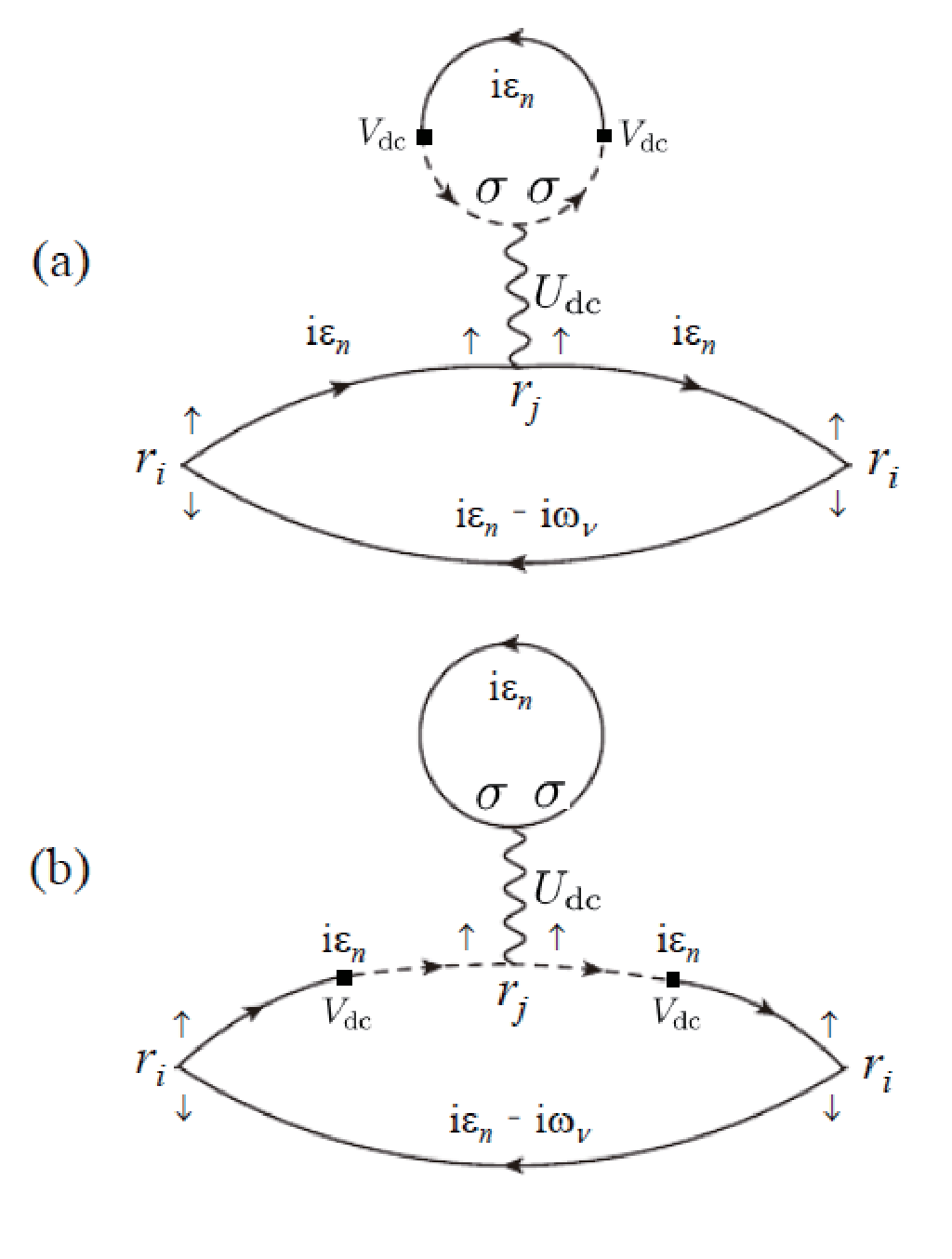}}
\caption{
Feynman diagram giving the NMR longitudinal relaxation rates $1/T_{1}T$ at Te (${\bf r}_{i}$) site 
in the lowest order with respect to the inter-orbital interaction $U_{\rm dc}$ 
which corresponds to a Hartree type self-energy correction. The summation with respect to spin 
component $\sigma=\uparrow$ and $\downarrow$. 
{he other diagrams are given by the inversion of upside down.}  
}
\label{Fig:Udc_H}
\end{center}
\end{figure}

{
Performing calculations similar to that obtaining the expression Eq.\ (\ref{Gamma1}) for 
${\rm Im}\Gamma_{\rm ph}^{\rm R}(\omega+{\rm i}\delta)$, the expression of 
${\rm Im}\Gamma_{\rm dcV}^{\rm R}(\omega+{\rm i}\delta)$ is 
given, to the leading order in $\omega$, as 
\begin{eqnarray}
{\rm Im}\Gamma_{\rm dcV}^{\rm R}(\omega+{\rm i}\delta)
=-\frac{4\pi\omega V_{\rm dc}^{2}{T}U_{\rm dc}}{\epsilon_{\rm d}^{2}}
\int_{-\infty}^{\infty}{\rm d}y\,
\frac{\partial}{\partial y}
\left({\rm th}\frac{y}{2T}\right)
\left[\rho({\bf r}_{ij},y)G_{\rm c}^{\prime{\rm R}}({\bf r}_{ij},y)\right]^{2}.
\label{rates7V}
\end{eqnarray}
Then, in the low temperature 
limit ($T\ll \epsilon_{\rm F}$), the 
${\rm Im}\Gamma_{\rm dcV}^{\rm R}(\omega+{\rm i}\delta)/\omega$ 
is reduced to a compact form as
\begin{equation}
\frac{\displaystyle {\rm Im}\Gamma_{\rm dcV}^{\rm R}(\omega+{\rm i}\delta)}{\omega}\approx
-\frac{8\pi V_{\rm dc}^{2}{T}U_{\rm dc}}{\epsilon_{\rm d}^{2}}
\left[\rho({\bf r}_{ij},0)G_{\rm c}^{\prime{\rm R}}({\bf r}_{ij},0)\right]^{2}.
\label{rates7KV}
\end{equation}
This term has the same form as Eq.\ (\ref{rates7D}) giving 
${\rm Im}\Gamma^{\rm R}_{\rm ph}(\omega+{\rm i}\delta)/\omega$ 
with $U_{\rm ph}$ being replaced by $-U_{\rm dc}$.  Therefore, it has an effect that 
$U_{\rm ph}$ in Eq.\ (\ref{rates7D}) is replaced by $(U_{\rm ph}-2U_{\rm dc})$.  
}

{
Similarly, the expression of 
${\rm Im}\Gamma_{\rm dcS}^{\rm R}(\omega+{\rm i}\delta)$ is 
given, to the leading order in $\omega$, as 
\begin{eqnarray}
{\rm Im}\Gamma_{\rm dcS}^{\rm R}(\omega+{\rm i}\delta)
=\frac{6\pi\omega V_{\rm dc}^{2}{T}U_{\rm dc}}{\epsilon_{\rm d}^{2}}
\int_{-\infty}^{\infty}{\rm d}y_{3}\,
\frac{\partial}{\partial y_{3}}
\left({\rm th}\frac{y_{3}}{2T}\right)
\left[\rho(0,y_{3})G_{\rm c}^{\prime{\rm R}}({\bf r}_{ij},y_{3})\right]^{2}.
\label{rates7J}
\end{eqnarray}
Then, in the low temperature 
limit ($T\ll \epsilon_{\rm F}$), the 
${\rm Im}\Gamma_{\rm dcS}^{\rm R}(\omega+{\rm i}\delta)/\omega$ [Eq.\ (\ref{rates7J})] 
is reduced to a compact form as
\begin{equation}
\frac{\displaystyle {\rm Im}\Gamma_{\rm dcS}^{\rm R}(\omega+{\rm i}\delta)}{\omega}\approx
\frac{12\pi V_{\rm dc}^{2}{T}U_{\rm dc}}{\epsilon_{\rm d}^{2}}
\left[\rho(0,0)G_{\rm c}^{\prime{\rm R}}({\bf r}_{ij},0)\right]^{2}.
\label{rates7KF}
\end{equation}

Substituting the expressions for $\rho(0,0)$ [Eq.\ (\ref{GF2A})] 
and $G_{\rm c}^{\prime{\rm R}}({\bf r}_{ij},0)$ [Eq.\ (\ref{GF1C})] and replacing the bare inter-orbital 
interaction $U^{0}_{\rm dc}$ by the renormalized one, $U_{\rm dc}(T)$ [Eq.\ (\ref{E9})], 
the relaxation rate $(1/T_{1}T)_{\rm dcS}$  [Eq.\ (\ref{rates0})] is given as  
\begin{equation}
\left(\frac{1}{T_{1}T}\right)_{\rm dcS} \approx A^{2}
\frac{{12}\pi (V_{\rm dc}N_{\rm F})^{2}}{\epsilon_{\rm d}^{2}}\,e^{-(r/\ell)}
\left[ {\frac{{\rm sin}(k_{\rm F}r)}{k_{\rm F}r}}J(k_{\rm F}r)\right]^{2}{T}U_{\rm dc}(T).
\label{rates7L}
\end{equation}
This formula is another central results of the present paper and offers us the basis for discussing 
anomalous behavior of the relaxation rate $1/T_{1}T$ in the region $T\sim T_{\rm K}$. }
\newpage

{
\subsection{Short summary for NMR relaxation rate}

The total relaxation rate $(1/T_{1}T)$ is given by the sum of $(1/T_{1}T)_{\rm ph}$ [Eq.\ (\ref{rates7G})], 
$(1/T_{1}T)_{\rm dcV}$ [Eq.\ (\ref{rates7KV})], and $(1/T_{1}T)_{\rm dcS}$ [Eq.\ \ref{rates7L})]  
as follows: 

\begin{eqnarray}
& &
\frac{1}{T_{1}T}\approx A^{{2}}
\frac{4\pi {N_{\rm F}^{2}}(V_{\rm dc}N_{\rm F})^{2}}
{\epsilon_{\rm d}^{2}}\,e^{-(r/\ell)}
\left[
\frac{\sin(k_{\rm F}r)}{k_{\rm F}r}\,J(k_{\rm F}r)\right]^{2}{T}\left[U_{\rm ph}(T)-2U_{\rm dc}(T)\right]
\nonumber
\\
& &
\qquad\quad
+A^{2}\frac{{12}\pi {N_{\rm F}^{2}}
(V_{\rm dc}N_{\rm F})^{2}}{\epsilon_{\rm d}^{2}}\,e^{-(r/\ell)}
\left[{\frac{{\rm sin}(k_{\rm F}r)}{k_{\rm F}r}}J(k_{\rm F}r)\right]^{2}{T}U_{\rm dc}(T),
\label{rates7M}
\end{eqnarray}
where we have used the expressions of $U_{\rm ph}(T)$ [Eq.\ (\ref{E8})] and 
$U_{\rm dc}(T)$ [Eq.\ (\ref{E9})]. 
Since $U_{\rm ph}$ and $U_{\rm dc}$ correspond to $J_{\perp}/2$ and $J_{z}/4$, respectively, 
as discussed in Appendices{\ref{equivalence}} and {\ref{Scaling}}, 
the ratio of $[U_{\rm ph}(T)-2U_{\rm dc}(T)]$ in the first term of 
Eq.\ (\ref{rates7M}) and $U_{\rm ph}(T)$ approaches zero toward $T=T_{\rm K}$ as decreasing 
temperature. Therefore, the first term { gives} 
less divergent behavior compared to the second term. 
}
{
On the other hand, the second term in Eq.\ (\ref{rates7M})
exhibits pronounced increase 
as $T$ decreases, toward $T=T_{\rm K}$ from the region $T\gsim T_{\rm K}$, through the $T$ 
dependence of $TU_{\rm dc}(T)$ (in a dimensionless form)  shown in Fig.\ \ref{Fig:Poorman} 
in which the $T$ dependence 
of $U_{\rm dc}(T)$ is given by the the one-loop order RG (or poorman's scaling) approximation 
as  
\begin{equation}
U_{\rm dc}(T)= \frac{1}{\displaystyle 4N_{\rm F}\log\frac{T}{T_{\rm K}}}.
\label{rates8}
\end{equation} 
 (See Eq.\ (\ref{E10}) for $y_{z}$ and definition of $U_{\rm dc}(E)\equiv y_{z}/4N_{\rm F}$ 
 in Appendix {\ref{Scaling}}.) 
Of course, the result of poorman's scaling ceases to be valid very near $T=T_{\rm K}$. 
Nevertheless, it would give an increasing tendency of $TU_{\rm dc}(T)$ around $T=T_{\rm K}$. 
The dotted line in Fig.\ \ref{Fig:Poorman} shows an expected $T$ dependence 
of $TU_{\rm dc}(T)$ at $T\lsim T_{\rm K}$, which is reasonable considering that 
the increasing tendency of $TU_{\rm dc}(T)$ already begins to appear at $T\simeq 2.7T_{\rm K}$, i.e.,  
from far higher temperature than $T_{\rm K}$, and that 
the divergent $T$ dependence in $U_{\rm dc}(T)$ at $T\ll T_{\rm K}$ works to suppress 
the Curie like divergence ($\propto 1/T$) of localized electron when entering into 
{the 
local Fermi liquid state~\cite{Nozieres} in which 
the Kondo-Yosida charge singlet state is formed} as in the case of magnetic Kondo 
problem.~\cite{ShibaKorringa} 
Since the {\it divergent} part in $1/T_{1}T$ [Eq.\ (\ref{rates7M})] is in proportion to 
$TU_{\rm dc}(T)$, this theoretical result for $1/T_{1}T$ qualitatively explains the anomalous temperature 
dependence of $1/T_{1}T$ observed  
in Pb$_{1-x}$Tl$_{x}$Te ($x\simeq 0.01$) reported in Ref.\ \citen{Mukuda}.
}
{
However, of course to obtain {more} quantitative result for the $T$ dependence in 
$1/T_{1}T$ at $T\lsim T_{\rm K}$, we need perform more solid calculations, such as numerical 
renormalization group method,~\cite{Matsuura} which is left for future study.  
}

\begin{figure}[h]
\begin{center}
\rotatebox{0}{\includegraphics[angle=0,width=0.6\linewidth]{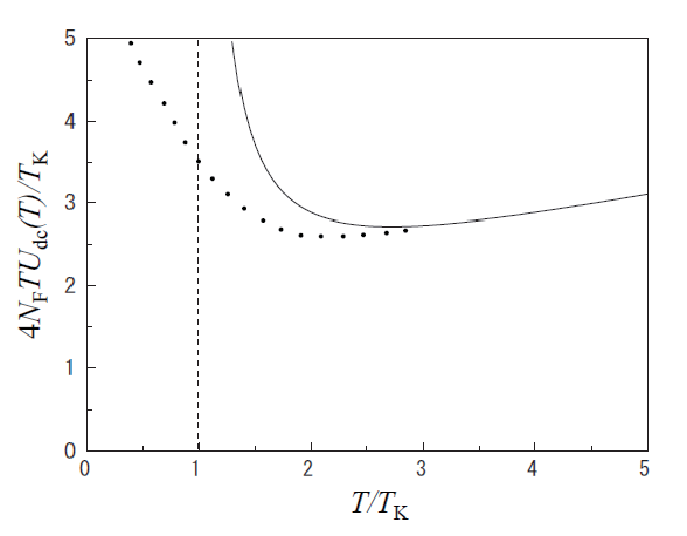}}
\caption{
{
$4N_{\rm F}TU_{\rm dc}(T)/T_{\rm K}$ vs $T/T_{\rm K}$ with $T_{\rm K}$ being the Kondo temperature 
in the one-loop order RG (poorman's scaling) approximation [Eq.\ (\ref{rates8})]. 
Dotted line is  {a guide to the eyes} 
for a qualitative behavior expected in exact treatment beyond 
poorman's scaling solution {as discussed in the text}. }
}
\label{Fig:Poorman}
\end{center}
\end{figure}

{
Concluding this section, it is remarked 
that the present relaxation mechanism is quite different from the case of magnetic Kondo 
impurity in which $1/T_{1}T$ is essentially in proportion to 
$J_{\perp}^{2}$ as discussed in Ref.\ \citen{muSR}.
This difference is traced back to the difference in the order of perturbation process giving the 
relaxation rates. 
In the present case, $1/T_{1}T$ is given by the first order process in the pair-hopping interaction 
$U_{\rm ph}$ and the inter-orbital interaction $U_{\rm dc}$, while that in the case of magnetic 
Kondo impurity is given by the second order process in 
the s-d exchange interaction $J_{\perp}$ causing the spin-flip process, as discussed in 
Ref.\ \citen{muSR}
}


{
\section{Anomaly of Knight Shift by Charge Kondo Effect} 
The temperature dependence of the Knight shift $K$ is related with ($1/T_{1}T$) 
through the Kramers-Kronig relation among the real and the imaginary part of  the {\it local} 
dynamical magnetic susceptibility $\chi(\omega+i\delta)$ at 
the site of the NMR measurement. This is because the Knight shift $K$ is proportional to the real part of 
$\chi(0)$ while ($1/T_{1}T$) is given by the imaginary part of  
$\chi(\omega+i\delta)/\omega$ by the Moriya formula [Eq. (\ref{rates0})] as 
\begin{eqnarray}
& &
\frac{1}{T_{1}T}
=A^{2}\left[\frac{1}{\omega}{\rm Im}\chi(\omega+{\rm i}\delta)\right]_{\omega=\omega_{\rm NMR}}
\nonumber
\\
& &
\qquad\,\,
\simeq
A^{2}\left[\frac{1}{\omega}{\rm Im}\chi(\omega+{\rm i}\delta)\right]_{\omega=0},
\label{I_1}
\end{eqnarray}
where $A$ is the hyper-fine coupling constant between electrons (quasiparticles) and nuclei,~\cite{Moriya} 
and  $\omega_{\rm NMR}$ is the frequency of NMR measurement. 
The explicit form of the Kramers-Kronig relation for  
$\chi(\omega+i\delta)\equiv \chi^{\prime}(\omega)+i\chi^{\prime\prime}(\omega)$ is 
\begin{eqnarray}
\chi^{\prime}(\omega)=\frac{1}{\pi}{\cal P}\int_{-\infty}^{\infty}d\omega^{\prime}\,
\frac{\chi^{\prime\prime}(\omega^{\prime})}{\omega^{\prime}-\omega}.
\label{I_2}
\end{eqnarray}
Therefore, the real part of $\chi^{\prime}(0)$ is given by the imaginary part of $\chi^{\prime\prime}(\omega)$ as 
\begin{eqnarray}
\chi^{\prime}(0)=\frac{1}{\pi}\int_{-\infty}^{\infty}d\omega^{\prime}\,
\frac{\chi^{\prime\prime}(\omega^{\prime})}{\omega^{\prime}}.
\label{I_3}
\end{eqnarray}
Here, we adopt the following parameterization: 
\begin{eqnarray}
\chi^{\prime\prime}(\omega)=\omega\left[\lim_{\omega\to0}\frac{\chi^{\prime\prime}(\omega)}{\omega}\right]
R(\omega).
\label{I4}
\end{eqnarray}
Note that the second factor $\lim_{\omega\to0}[\chi^{\prime\prime}(\omega)/\omega]$, or ($1/T_{1}T$),  
exhibits increasing 
tendency as the temperature decreases across the Kondo temperature $T_{\rm K}$ as discussed in subsection 3.3.  
{
On the other hand, the remnant factor $R(\omega)$ is a positive, even and decreasing function in $\omega$,  
with lim$_{\omega\to 0}R(\omega)=1$ and $\lim_{|\omega|\to \infty}R(\omega)\propto \omega^{-2}$
at least.~\cite{AGD2} 
Otherwise, it exhibits a moderate 
$\omega$- and $T$-dependence in entire $\omega$ region.  }
As a result, the correction to the Knight shift $\Delta K$, given with the use of 
$\Delta \chi^{\prime}(0)$ [cf. Eq. (\ref{I_3})], exhibits the anomaly similar 
to the correction to the NMR relaxation rate $\Delta(1/T_{1}T)$.  
This is because the following relation holds.  
\begin{eqnarray}
\Delta K=A\Delta\chi^{\prime}(0)\simeq\Delta\left(\frac{1}{T_{1}T}\right)
\frac{1}{\pi A}\int_{-\infty}^{\infty}d\omega^{\prime}R(\omega^{\prime}), 
\label{I5}
\end{eqnarray}
where the integration of $R(\omega^{\prime})$ is expected to have only weak $T$-dependence, 
{
because the most dominant $T$-dependence in $\chi^{\prime\prime}(\omega)$ has been taken into account 
by the second factor in Eq. (\ref{I4}).
}  

Indeed, such $T$ increasing tendency of the Knight shift around $T_{\rm K}$ has been recently observed 
by experiment.~\cite{Mukuda2} 
This implies that the Koriinga relation $1/T_{1}T\propto K^{2}$ is apparently broken in general. 
Here, it is crucial that the present subject is the {\it local} 
impurity problem which is free from the long range magnetic correlations due to, e.g., the enhanced antiferromagenetic 
fluctuations as observed in the case of high T$_{\rm c}$ cuprates.~\cite{High-Tc}}

The aspect of the Knight shift discussed here has been verified by explicit microscopic calculations similar to 
obtaining the anomalous  behavior of the NMR relaxation rate $1/T_{1}$, which will be published elsewhere.

\section{Summary}
We have shown that the anomalous NMR response observed in Pb$_{1-x}$Tl$_{x}$Te ($x\sim 0.01$) 
can be explained by the charge Kondo  effect which is caused by the pair-hopping interaction 
$U_{\rm ph}$ and the inter-orbital interaction $U_{\rm dc}$ between localized orbital on Tl and 
conduction electrons doped in the semiconductor PbTe.  Sharp increase observed in the NMR 
relaxation rate $1/T_{1}T$ of $^{125}$Te at $T<10$K can be understood essentially as the increase of 
$U_{\rm ph}$ {and $U_{\rm dc}$} 
due to Kondo-like renormalization in the region $T\gsim T_{\rm K}$ 
because $U_{\rm ph}$ and $U_{\rm dc}$ can mediate the spin-flip of conduction electrons as shown in 
Fig.\ \ref{Fig:1}, and Figs.\ \ref{Fig:Udc_V}, \ref{Fig:Udc_F}, and \ref{Fig:Udc_H}, respectively .
We have also shown that the Knight shift K is also enhanced in proportion to 
the relaxation rate $1/T_{1}T$ in the region of temperature 
where $1/T_{1}T$ is enhanced. 
In this sense, the Korringa relation is apparently broken in the system with the charge Kondo effect.



\section*{Acknowledgments}
We are grateful to H. Mukuda for stimulating discussions on the experimental results of $^{125}$Te 
NMR in Pb$_{1-x}$Tl$_x$Te (x=0.01). 
This work is supported by the Grant-in-Aid for Scientific Research 
{(Nos. 15K17694 and 17K05555)} 
from the Japan Society for the Promotion of Science. 

\newpage
\appendix
\section{
Equivalence of Pair-Hopping and Inter-orbital Interactions to Pseudo-Spin Exchange 
Interactions}\label{equivalence}
Here we discuss why the pair-hopping interaction $U_{\rm ph}$ is enhanced in the scattering channel 
${\rm i}\varepsilon_{n}\ \to -{\rm i}\varepsilon_{n}$, shown in Figs.\ \ref{Fig:A1} and \ref{Fig:A3}. 
The reason why it is enhanced by the charge Kondo effect is understood intuitively by 
the fact that the pair-hopping interaction is mapped to 
that of spin-flipping interaction, i.e., the heart of the Kondo interaction, 
by the canonical transformation for both the localized electron d and conduction electrons 
with $\downarrow$ spin and $\uparrow$ spin as 
\begin{eqnarray}
d^{\dagger}_{\downarrow}\ \to\  {\tilde d}_{\downarrow}\ &\hbox{and}&\ 
c^{\dagger}_{{\bf k}\downarrow}\ \to\  {\tilde c}_{{\bf k}\downarrow},
\label{Shiba1}
\\
d^{\dagger}_{\uparrow}\ \to\  {\tilde d}^{\dagger}_{\uparrow}\ &\hbox{and}&\ 
c^{\dagger}_{{\bf k}\uparrow}\ \to\  {\tilde c}^{\dagger}_{{\bf k}\uparrow}.  
\label{Shiba2}
\end{eqnarray}
This is a variant of the canonical transformation introduced by Shiba.~\cite{Shiba,Micnas}  
Namely, by the transformations [Eqs.\ (\ref{Shiba1}) and (\ref{Shiba2})], 
the pair-hopping interaction is mapped as follows: 
\begin{eqnarray}
& &
U_{\rm ph}\sum_{{\bf k},{\bf k}^{\prime}}
\left(d^{\dagger}_{\uparrow}d^{\dagger}_{\downarrow}
c_{{\bf k}\downarrow}c_{{\bf k}^{\prime}\uparrow}+{\rm h.c.}\right)
\nonumber
\\
& &
\qquad
\to\ 
U_{\rm ph}\sum_{{\bf k},{\bf k}^{\prime}}
\left({\tilde d}^{\dagger}_{\uparrow}{\tilde d }_{\downarrow}
{\tilde c}^{\dagger}_{{\bf k}\downarrow}{\tilde c}_{{\bf k}^{\prime}\uparrow}+{\rm h.c.}
\right)
\equiv U_{\rm ph}\left({\tilde S}_{\rm d}^{+}{\tilde S}_{{\bf k},{\bf k}^{\prime}}^{-}+{\rm h.c.}
\right).
\label{mapping}
\end{eqnarray}
Therefore, the pair-hopping interaction $U_{\rm ph}$, which is equivalent to $J_{\perp}/2$ in the 
anisotropic s-d model{~\cite{PoormanScaling}} is enhanced by the Kondo effect 
in the mapped world. The spin-flipping exchange interaction $U_{\rm ph}$ in the mapped world is 
represented by the Feynman diagram shown in Fig.\ \ref{Fig:A1}(a), while the pair-hopping interaction 
$U_{\rm ph}$ in the original world is given by the Feynman diagram shown in Fig.\ \ref{Fig:A1}(b). 
Note that the Matsubara frequency of the conduction electrons with $\downarrow$ spin 
has the opposite sign of that of the $\uparrow$ spin because the direction of propagation 
in the imaginary time is opposite. 
Namely, the elastic scattering with ${\rm i}\varepsilon_{n}\ \to\, {\rm i}\varepsilon_{n}$ 
in the mapped world causing the Kondo effect  corresponds to the scattering with 
 ${\rm i}\varepsilon_{n}\ \to\, -{\rm i}\varepsilon_{n}$ in the original world. 
This is the reason why the process shown in Fig.\ \ref{Fig:1} is selectively enhanced. 

\begin{figure}[h]
\begin{center}
\rotatebox{0}{\includegraphics[angle=0,width=0.7\linewidth]{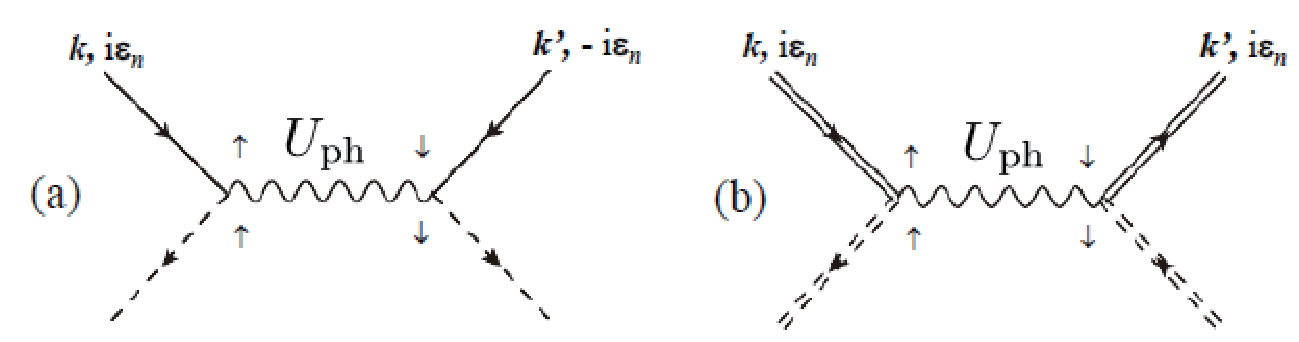}}
\caption{(a) Feynman diagram representing the pair-hopping process in the original world, 
and (b) Feynman diagram representing the spin-flipping exchange process in the mapped world 
by the canonical transformation [Eqs.\ (\ref{Shiba1}) and (\ref{Shiba2})].  Wavy line represents 
the pair-hopping interaction $U_{\rm ph}$, lines with arrow represent the 
Green function of conduction electrons in the original world, and double lines represent that in the 
mapped world. Dashed lines with arrow denote the Green functions of the localized electron d both 
in original and mapped worlds. }
\label{Fig:A1}
\end{center}
\end{figure}

By the transformation [Eqs.\ (\ref{Shiba1}) and (\ref{Shiba2})], the spin dependent density of states (DOS), 
$D_{\sigma}(\varepsilon)$, of conduction electrons and localized d electron change from that shown in 
Fig. \ref{Fig:A2}(a) to 
that in Fig. \ref{Fig:A2}(b). Namely, symmetry with respect to $\uparrow$ and $\downarrow$ spins 
is broken. Nevertheless, the Kondo effect is possible if the finite DOS of conduction electrons 
remain at the Fermi level and the energy level of localized electron with 
$\uparrow$ and $\downarrow$ spins are degenerate, i.e., 
$\epsilon_{\rm d}=-\epsilon_{\rm d}-U_{\rm dc}$, 
as shown in Fig.\ \ref{Fig:A2}.  The latter condition is satisfied in the negative-$U$ Anderson model 
for rather wide doping rates of negative-$U$ ions as discussed in ref.\ \citen{Taraphder}. 
It was also shown by the present authors~\cite{Matsuura} that the negative-$U$ effect is realized 
in the model described by the Hamiltonian [Eq.\ (\ref{effmod})]. 
In this sense, it is assured that the condition for zero 
magnetic field on the localized electron is satisfied in a self-consistent fashion.  

\begin{figure}[h]
\begin{center}
\rotatebox{0}{\includegraphics[angle=0,width=0.7\linewidth]{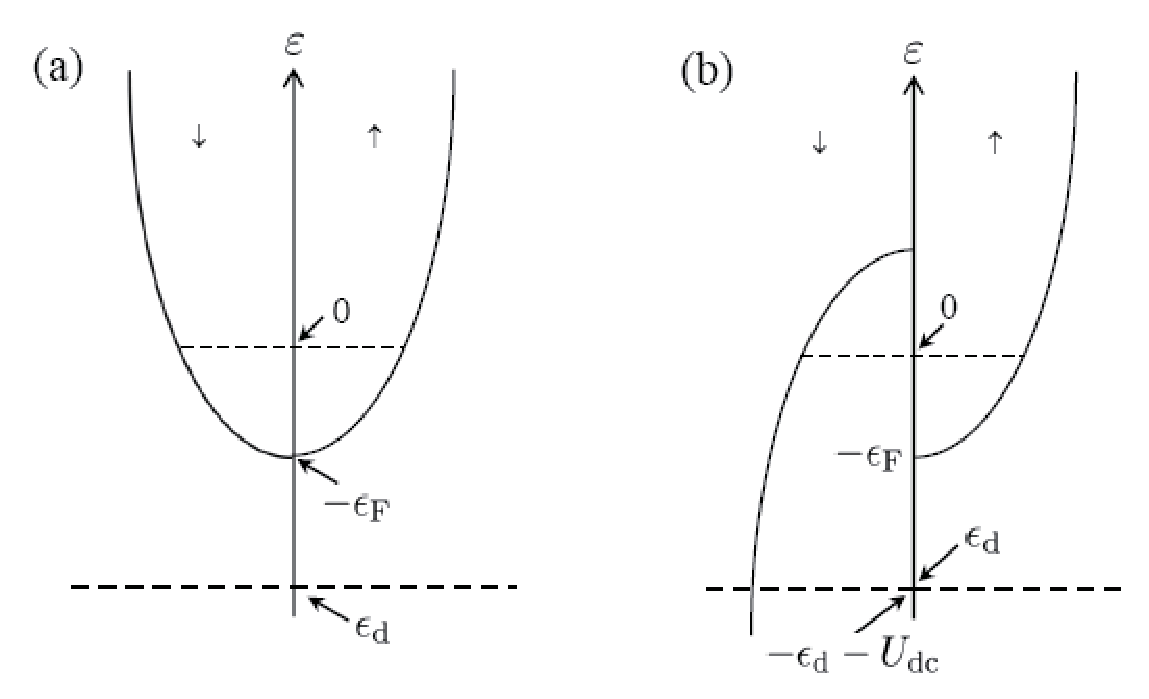}}
\caption{
Spin dependent DOS, $D_{\sigma}(\varepsilon)$: (a)  in the original world, 
and (b) in the transformed world by the transformation Eqs.\ Eqs.\ (\ref{Shiba1}) and (\ref{Shiba2}). 
Note that the origin of energy is $\epsilon_{\rm F}$, or energy is measured from $\epsilon_{\rm F}$.
}
\label{Fig:A2}
\end{center}
\end{figure}

Similarly, the inter-orbital interaction $U_{\rm dc}$ is also enhanced by the charge Kondo effect. 
Indeed, the inter-orbital interaction is mapped by the transformations [Eqs.\ (\ref{Shiba1}) and 
(\ref{Shiba2})] as follows: 
\begin{eqnarray}
& &
U_{\rm dc}\sum_{{\bf k},{\bf k}^{\prime}}\sum_{\sigma\sigma^{\prime}}
d^{\dagger}_{\sigma}d_{\sigma}
c_{{\bf k}\sigma^{\prime}}c_{{\bf k}^{\prime}\sigma^{\prime}}
\nonumber
\\
& &
\qquad
\to\ 
U_{\rm dc}\sum_{{\bf k},{\bf k}^{\prime}}
\left({\tilde d}^{\dagger}_{\uparrow}{\tilde d }_{\uparrow}
-
{\tilde d}^{\dagger}_{\downarrow}{\tilde d }_{\downarrow}\right)
\left(
{\tilde c}^{\dagger}_{{\bf k}\uparrow}{\tilde c}_{{\bf k}^{\prime}\uparrow}
-
{\tilde c}^{\dagger}_{{\bf k}\downarrow}{\tilde c}_{{\bf k}^{\prime}\downarrow}\right)
\equiv 4U_{\rm dc}{\tilde S}_{\rm d}^{z}{\tilde S}_{{\bf k},{\bf k}^{\prime}}^{z}.
\label{mapping2}
\end{eqnarray}
Therefore, the inter-orbital interaction $U_{\rm dc}$ is enhanced by the Kondo effect 
in the mapped world because it corresponds to $J_{z}/4$ in the anisotropic s-d 
model.~\cite{Anderson}  In this sense, the pair-hopping interaction $U_{\rm ph}$ and 
inter-orbital interaction $U_{\rm dc}$ should be treated impartially as in the case of 
magnetic Kondo effect.~\cite{Anderson} 

\section{Spectral Function of Conduction Electrons}\label{SpectralFunction}
In this Appendix, we derive the spectral function, Eq.\ (\ref{GF2}), for the conduction electrons. 
First, we note that the Green function $G_{\rm c}({\bf r},{\rm i}\varepsilon_{n})$ of conduction 
electrons with impurity scattering is given by    
\begin{equation}
G_{\rm c}({\bf r},{\rm i}\varepsilon_{n})=e^{-(r/2\ell)}G_{\rm c}^{(0)}({\bf r},{\rm i}\varepsilon_{n}), 
\label{B1}
\end{equation}
where $\ell$ is the mean-free path of the impurity scattering, and $G_{\rm c}^{(0)}({\bf r},{\rm i}\varepsilon_{n})$ 
is the Green function in the pure system without impurity scattering.~\cite{AGD}  
An explicit form of $G_{\rm c}^{(0)}({\bf r},{\rm i}\varepsilon_{n})$ is calculated as follows:
\begin{eqnarray}
& &
G_{\rm c}^{(0)}({\bf r},{\rm i}\varepsilon_{n})\,={\frac{1}{N}}\int \displaystyle \frac{{\rm d} {\bf k}}{(2\pi)^3} \, 
\frac{e^{{\rm i}{\bf k} \cdot {\bf r}}}{{\rm i}\epsilon_n -\xi_k}
\nonumber
\\
& &
\qquad\qquad\,\,\,\,
=\frac{1}{2\pi^{2}{N}}\frac{1}{r}\int_{0}^{k_{\rm c}}{\rm d}k\,k\sin(kr)\,\frac{1}{{\rm i}\varepsilon_{n}-\xi_{k}},
\label{B2}
\end{eqnarray}
where {the system volume is taken as the unit of volume, $N$ is the number of lattice sites, and}  $\xi_{k}\equiv (k^{2}/2m)-\mu$, and we have introduced the upper cut-off wave number $k_{\rm c}$.  
Then, the imaginary part of the retarded function 
${\rm Im}G_{\rm c}^{(0){\rm R}}({\bf r},\varepsilon+{\rm i}\delta)$ is calculated as follows: 
\begin{eqnarray}
& &
{\rm Im}G_{\rm c}^{(0){\rm R}}({\bf r},\varepsilon+{\rm i}\delta)\,=
-\frac{1}{2\pi{N}}\frac{1}{r}\int_{0}^{k_{\rm c}}{\rm d}k\,k\sin(kr)\,\delta(\varepsilon-\xi_{k})
\nonumber
\\
& &
\qquad\qquad\qquad\quad\,\,\,
=-\frac{m}{2\pi{N}}\frac{1}{r}\int_{-\epsilon_{\rm F}}^{\epsilon_{\rm c}}{\rm d}\xi\,
\sin\left[\sqrt{\frac{2m\xi}{k_{\rm F}^{2}}+1}\,(k_{\rm F}r)\right]\,\delta(\varepsilon-\xi)
\nonumber
\\
& &
\qquad\qquad\qquad\quad\,\,\,
=-\frac{m}{2\pi{N}}\frac{1}{r}
\sin\left[\sqrt{\frac{\varepsilon}{\epsilon_{\rm F}}+1}\,(k_{\rm F}r)\right]\,
\theta(\varepsilon+\epsilon_{\rm F})\theta(\varepsilon-\epsilon_{\rm c}),
\label{B3}
\end{eqnarray}
where {
we have approximated $\mu$ by $-\epsilon_{\rm F}$ because we are interested in the low 
temperature region $T\ll \epsilon_{\rm F}$} 
and $\epsilon_{\rm c}$ is the upper 
cut-off energy corresponding to $k_{\rm c}$. Therefore, the spectral  function 
$\rho({\bf r},\varepsilon)\equiv -(1/\pi){\rm Im}G_{\rm c}^{(0){\rm R}}({\bf r},\varepsilon+{\rm i}\delta)$ is 
given by Eq.\ (\ref{GF2}).

\section{Calculation of $\Gamma_{\rm ph}^{\rm R}(\omega+{\rm i}\delta)$}\label{Gamma}
In this Appendix, we calculate the expression [Eq.\ (\ref{rate1})] and derive the expression of 
Im$\Gamma_{\rm ph}^{\rm R}$ [Eq.\ (\ref{Gamma1})].  
As is justified in Appendix\ref{DirectOverlap}, $G_{\rm d}(-{\rm i}\varepsilon_{n}+{\rm i}\omega_{\nu})
G_{\rm d}({\rm i}\varepsilon_{n})$ in Eq.\ (\ref{rate1}) can be approximated by $1/\epsilon_{\rm d}^{2}$. 
Then, ${\tilde \Gamma}_{\rm ph}({\rm i}\omega_{\nu})\equiv \Gamma_{\rm ph}({\rm i}\omega_{\nu})/T$ 
[see Eq.\ (\ref{rate1})] is given as 
\begin{eqnarray}
{\tilde \Gamma}_{\rm ph}({\rm i}\omega_{\nu})
= \frac{2V_{\rm dc}^{2}U_{\rm ph}}{\epsilon_{\rm d}^{2}}T\sum_{\varepsilon_{n}}
\left(\prod_{\ell=1}^{4}\int_{-\infty}^{\infty}{\rm d}y_{\ell}\right)\,
\frac{\rho({\bf r}_{ij},y_{1})}{{\rm i}\varepsilon_{n}-y_{1}}\,
\frac{\rho({\bf r}_{ij},y_{2})}{{\rm i}\varepsilon_{n}+{\rm i}\omega_{\nu}-y_{2}}\,
\frac{\rho({\bf r}_{ij},y_{3})}{-{\rm i}\varepsilon_{n}-y_{3}}\,
\frac{\rho({\bf r}_{ij},y_{4})}{-{\rm i}\varepsilon_{n}+{\rm i}\omega_{\nu}-y_{4}}.
\label{rates3}
\end{eqnarray}
The summation with respect to $\varepsilon_{n}$ is performed in a standard way by transforming 
the summation to the integration along the axes Im$z=0$ and Im$z=\pm\omega_{\nu}$ on $z$-plane, 
where one is just above these axes and another is just below in the counter direction. The result 
along  Im$z=0$, $\Gamma_{\rm I}({\rm i}\omega_{\nu})$, is given by
\begin{eqnarray}
& & 
\Gamma_{\rm I}({\rm i}\omega_{\nu})
=\frac{2V_{\rm dc}^{2}U_{\rm ph}}{\epsilon_{\rm d}^{2}}
\left(\prod_{\ell=1}^{4}\int_{-\infty}^{\infty}{\rm d}y_{\ell}\right)\,
\rho({\bf r}_{ij},y_{1})\rho({\bf r}_{ij},y_{2})\rho({\bf r}_{ij},y_{3})\rho({\bf r}_{ij},y_{4})
\nonumber
\\
& &
\qquad\qquad\qquad\qquad
\times
{\rm th}\left(\frac{y_{1}}{2T}\right)\,\frac{1}{y_{1}+y_{3}}\,
\frac{1}{y_{1}-y_{2}+{\rm i}\omega_{\nu}}\,
\frac{1}{-y_{1}-y_{4}+{\rm i}\omega_{\nu}},
\label{rates4I}
\end{eqnarray}
and those along Im$z=\pm\omega_{\nu}$, $\Gamma_{\rm II}({\rm i}\omega_{\nu})$, are both given by 
\begin{eqnarray}
& & 
\Gamma_{\rm II}({\rm i}\omega_{\nu})
=-\frac{V_{\rm dc}^{2}U_{\rm ph}}{\epsilon_{\rm d}^{2}}
\left(\prod_{\ell=1}^{4}\int_{-\infty}^{\infty}{\rm d}y_{\ell}\right)\,
\rho({\bf r}_{ij},y_{1})\rho({\bf r}_{ij},y_{2})\rho({\bf r}_{ij},y_{3})\rho({\bf r}_{ij},y_{4})
\nonumber
\\
& &
\qquad\qquad\qquad\qquad
\times
{\rm th}\left(\frac{y_{2}}{2T}\right)\,\frac{1}{y_{2}-y_{1}-{\rm i}\omega_{\nu}}\,
\frac{1}{-y_{2}-y_{3}+{\rm i}\omega_{\nu}}\,
\frac{1}{-y_{2}-y_{4}+2{\rm i}\omega_{\nu}}. 
\label{rates4II}
\end{eqnarray}

\noindent
After analytic continuation, ${\rm i}\omega_{\nu}\to \omega+{\rm i}\delta$ in Eq.\ (\ref{rates4I}), 
and taking an imaginary part, we obtain
\begin{eqnarray}
& & 
{\rm Im}\Gamma_{\rm I}^{\rm R}(\omega+{\rm i}\delta)
=\frac{2\pi V_{\rm dc}^{2}U_{\rm ph}}{\epsilon_{\rm d}^{2}}
\left(\prod_{\ell=1}^{4}\int_{-\infty}^{\infty}{\rm d}y_{\ell}\right)\,
\rho({\bf r}_{ij},y_{1})\rho({\bf r}_{ij},y_{2})\rho({\bf r}_{ij},y_{3})\rho({\bf r}_{ij},y_{4})
\nonumber
\\
& &
\qquad\qquad
\times{\rm th}\left(\frac{y_{4}-\omega}{2T}\right)
\frac{1}{y_{2}+y_{4}-2\omega}\left[
\frac{\delta(y_{1}+y_{4}-\omega)}{-y_{3}+y_{4}-\omega}
+\frac{\delta(y_{1}-y_{4}+\omega)}{y_{3}+y_{4}-\omega}
\right],
\label{rates5I}
\end{eqnarray}
where and hereafter the integration implies the principal value integration. 
In deriving E.\ (\ref{rates5I}), we have used the property that $\rho({\bf r}_{ij},y_{2})\rho({\bf r}_{ij},y_{4})$ is symmetric with respect to interchange $y_{2}\rightleftharpoons y_{4}$. 
Similarly, for Eq.\ (\ref{rates4II}), we obtain 
\begin{eqnarray}
& & 
{\rm Im}\Gamma_{\rm II}^{\rm R}(\omega+{\rm i}\delta)
=-\frac{\pi V_{\rm dc}^{2}U_{\rm ph}}{\epsilon_{\rm d}^{2}}
\left(\prod_{\ell=1}^{4}\int_{-\infty}^{\infty}{\rm d}y_{\ell}\right)\,
\rho({\bf r}_{ij},y_{1})\rho({\bf r}_{ij},y_{2})\rho({\bf r}_{ij},y_{3})\rho({\bf r}_{ij},y_{4})
\nonumber
\\
& &
\qquad\qquad\qquad\qquad
\times{\rm th}\left(\frac{y_{2}}{2T}\right)\Biggl\{
\frac{1}{y_{2}+y_{4}-2\omega}\left[
\frac{\delta(y_{2}-y_{1}-\omega)}{y_{2}+y_{3}-\omega}
-\frac{\delta(y_{2}+y_{3}-\omega)}{y_{1}-y_{2}+\omega}
\right]
\nonumber
\\
& &
\qquad\qquad\qquad\qquad\qquad\qquad\qquad\qquad
-\frac{\delta(y_{2}+y_{4}-2\omega)}{{(y_{1}-y_{2}+\omega)(y_{2}+y_{3}-\omega)}}\Biggr\}.
\label{rates5II}
\end{eqnarray}

Performing the integration with respect to $y_{1}$, Eq.\ (\ref{rates5I}) is reduced to 
\begin{eqnarray}
& &
{\rm Im}\Gamma_{\rm I}^{\rm R}(\omega+{\rm i}\delta)=
\frac{2\pi V_{\rm dc}^{2}U_{\rm ph}}{\epsilon_{\rm d}^{2}}
\left(\prod_{\ell=2}^{4}\int_{-\infty}^{\infty}{\rm d}y_{\ell}\right)\,
{\rm th}\left(\frac{y_{4}-\omega}{2T}\right)
\frac{\rho({\bf r}_{ij},y_{2})\rho({\bf r}_{ij},y_{3})\rho({\bf r}_{ij},y_{4})}{y_{2}+y_{4}-2\omega}
\nonumber
\\
& &\qquad\qquad\qquad\qquad\qquad\qquad\qquad\qquad
\times
\left[\frac{\rho({\bf r}_{ij},y_{4}-\omega)}{y_{3}+y_{4}-\omega}
-\frac{\rho({\bf r}_{ij},\omega-y_{4})}{y_{3}-y_{4}+\omega}
\right].
\label{rates6I}
\end{eqnarray}
Similarly, after performing the integration with respect to $y_{2}$, Eq.\ (\ref{rates5II}) is reduced to 
\begin{eqnarray}
& &
{\rm Im}\Gamma_{\rm II}^{\rm R}(\omega+{\rm i}\delta)=
-\frac{\pi V_{\rm dc}^{2}U_{\rm ph}}{\epsilon_{\rm d}^{2}}
\left(\prod_{\ell=1,3,4}\int_{-\infty}^{\infty}{\rm d}y_{\ell}\right)\,
\rho({\bf r}_{ij},y_{1})\rho({\bf r}_{ij},y_{3})\rho({\bf r}_{ij},y_{4})
\nonumber 
\\
& &
\times
\Biggl[
{\rm th}\left(\frac{{y_{1}}}{2T}\right)
\frac{\rho({\bf r}_{ij},y_{1}+\omega)}{(-y_{1}-y_{3})(-y_{1}-y_{4}+\omega)}
-{\rm th}\left(\frac{{y_{3}}}{2T}\right)
\frac{\rho({\bf r}_{ij},{y_{3}})}{(-y_{1}-y_{3})(y_{3}-y_{4}+\omega)}
\nonumber
\\
& &\qquad\qquad\qquad\qquad\qquad
-{\rm th}\left(\frac{-y_{4}+2\omega}{2T}\right)
\frac{\rho({\bf r}_{ij},-y_{4}+2\omega)}{(-y_{1}-y_{4}+\omega)(-y_{3}+y_{4}-\omega)}
\Biggr].
\label{rates6II}
\end{eqnarray}

By changing the integration variable from 
$y_{4}$ to $y_{4}-\omega$, ${\rm Im}\Gamma_{\rm I}^{\rm R}(\omega+{\rm i}\delta)$ 
[E\ (\ref{rates6I})] is simplified as 
\begin{eqnarray}
& &
{\rm Im}\Gamma_{\rm I}^{\rm R}(\omega+{\rm i}\delta)=
 \frac{2\pi V_{\rm dc}^{2}U_{\rm ph}}{\epsilon_{\rm d}^{2}}
\left(\prod_{\ell=2}^{4}\int_{-\infty}^{\infty}{\rm d}y_{\ell}\right)\,
{\rm th}\left(\frac{y_{4}}{2T}\right)
\frac{\rho({\bf r}_{ij},y_{2})\rho({\bf r}_{ij},y_{3})\rho({\bf r}_{ij},y_{4}+\omega)}{y_{2}+y_{4}-\omega}
\nonumber
\\
& &\qquad\qquad\qquad\qquad\qquad\qquad\qquad\qquad
\times
\left[\frac{\rho({\bf r}_{ij},y_{4})}{y_{3}+y_{4}}
-\frac{\rho({\bf r}_{ij},-y_{4})}{y_{3}-y_{4}}
\right].
\label{rates7I}
\end{eqnarray}
Similarly, by changing the integration variables from $y_{1}+\omega$ to $y_{1}$ 
{in the first term, and 
from $-y_{3}+\omega$ to $y_{3}$ and interchanging $y_{1}\rightleftharpoons y_{3}$ 
in} the second term of Eq.\ (\ref{rates6II}), 
${\rm Im}\Gamma_{\rm II}^{\rm R}(\omega+{\rm i}\delta)$ [E\ (\ref{rates6II})] is simplified as
\begin{eqnarray}
& &
{\rm Im}\Gamma_{\rm II}^{\rm R}(\omega+{\rm i}\delta)=
-\frac{\pi V_{\rm dc}^{2}U_{\rm ph}}{\epsilon_{\rm d}^{2}}
\left(\prod_{\ell=1,3,4}\int_{-\infty}^{\infty}{\rm d}y_{\ell}\right)\,
\Biggl\{
{\rm th}\left(\frac{y_1}{2T}\right)\Biggl[
\frac{\rho({\bf r}_{ij},y_{1}-\omega)\rho({\bf r}_{ij},y_{1})\rho({\bf r}_{ij},y_{3})\rho({\bf r}_{ij},y_{4})}
{(-y_{1}-y_{3}+\omega)(-y_{1}-y_{4}+2\omega)}
\nonumber
\\
& &\qquad\qquad\qquad\qquad\qquad\qquad\qquad\qquad\qquad\quad
+\frac{\rho({\bf r}_{ij},y_{1}+\omega)\rho({\bf r}_{ij},-y_{1})\rho({\bf r}_{ij},y_{3})\rho({\bf r}_{ij},y_{4})}
{(-y_{1}-y_{3}{-}\omega)(y_{1}-y_{4}+2\omega)}
\Biggr]
\nonumber
\\
& &\qquad\qquad\qquad\qquad
-{\rm th}\left(\frac{-y_{4}+\omega}{2T}\right)
\frac{\rho({\bf r}_{ij},y_{1})\rho({\bf r}_{ij},y_{3})\rho({\bf r}_{ij},y_{4}+\omega)\rho({\bf r}_{ij},-y_{4}+\omega)}
{(y_{1}+y_{4})(y_{3}-y_{4})}
\Biggr\}.
\label{rates7II}
\end{eqnarray}
To perform the integration with respect to $y_{2}$ and $y_{3}$ in the first term in the brace of 
Eq.\ (\ref{rates7I}), we use the spectral representation, 
Eq.\ (\ref{GF1}), for the Green function $G_{\rm c}$. Namely, the real part of the retarded 
Green function of 
conduction electrons, $G_{\rm c}^{\prime{\rm R}}({\bf r},\varepsilon)$, is given by 
\begin{equation}
G_{\rm c}^{\prime{\rm R}}({\bf r},\varepsilon)=\int_{-\infty}^{\infty}{\rm d}y\,
\frac{\rho({\bf r},y)}{\varepsilon-y}.
\label{GF1A}
\end{equation}
With the use of this relation, Eq.\ (\ref{rates7I}) is transformed to more compact form as 
\begin{eqnarray}
& &
{\rm Im}\Gamma_{\rm I}^{\rm R}(\omega+{\rm i}\delta)=
 \frac{2\pi V_{\rm dc}^{2}U_{\rm ph}}{\epsilon_{\rm d}^{2}}
\int_{-\infty}^{\infty}{\rm d}y_{4}\,
{\rm th}\left(\frac{y_{4}}{2T}\right)
\rho({\bf r}_{ij},y_{4}+\omega)G_{\rm c}^{\prime{\rm R}}({\bf r}_{ij},-y_{4}+\omega)
\nonumber
\\
& &\qquad\qquad\qquad\qquad
\times
\left[
\rho({\bf r}_{ij},y_{4})G_{\rm c}^{\prime{\rm R}}({\bf r}_{ij},-y_{4})
-\rho({\bf r}_{ij},-y_{4})G_{\rm c}^{\prime{\rm R}}({\bf r}_{ij},y_{4})
\right].
\label{rates7A}
\end{eqnarray}
Similarly, Eq.\ (\ref{rates7II}) is transformed to the following from  
\begin{eqnarray}
& &
{\rm Im}\Gamma_{\rm II}^{\rm R}(\omega+{\rm i}\delta)=
-\frac{\pi V_{\rm dc}^{2}U_{\rm ph}}{\epsilon_{\rm d}^{2}}
\Biggl\{
\int_{-\infty}^{\infty}{\rm d}y_{1}\,
{\rm th}\left(\frac{y_{1}}{2T}\right)
\nonumber
\\
& &\qquad\qquad\qquad\qquad\qquad
\biggl[
\rho({\bf r}_{ij},y_{1}-\omega)\rho({\bf r}_{ij},y_{1})G_{\rm c}^{\prime{\rm R}}({\bf r}_{ij},-y_{1}+\omega)
G_{\rm c}^{\prime{\rm R}}({\bf r}_{ij},-y_{1}+2\omega)
\nonumber
\\
& &
\qquad\qquad\qquad\qquad\qquad\quad
+\rho({\bf r}_{ij},y_{1}+\omega)\rho({\bf r}_{ij},-y_{1})G_{\rm c}^{\prime{\rm R}}({\bf r}_{ij},-y_{1}-\omega)
G_{\rm c}^{\prime{\rm R}}({\bf r}_{ij},y_{1}+2\omega)
\biggr]
\nonumber
\\
& &\qquad\quad
+\int_{-\infty}^{\infty}{\rm d}y_{1}\,
{\rm th}\left(\frac{y_{4}-\omega}{2T}\right)
\left[\rho({\bf r}_{ij},y_{4}+\omega)\rho({\bf r}_{ij},-y_{4}+\omega)
G_{\rm c}^{\prime{\rm R}}({\bf r}_{ij},-y_{4})G_{\rm c}^{\prime{\rm R}}({\bf r}_{ij},y_{4})
\right]
\Biggr\}.
\label{rates7B}
\end{eqnarray}
Changing the integration variables from $y_{1}$ to $y_{1}-\omega$ and from $y_{1}$ to $-y_{1}+\omega$ 
in the first and second term in the bracket of Eq.\ (\ref{rates7B}),  
the first term in the brace of Eq.\ (\ref{rates7B}) is transformed to 
\begin{eqnarray}
& &
-\frac{\pi V_{\rm dc}^{2}U_{\rm ph}}{\epsilon_{\rm d}^{2}}
\int_{-\infty}^{\infty}{\rm d}y_{1}\,
{\rm th}\left(\frac{y_{1}+\omega}{2T}\right)
\biggl[
\rho({\bf r}_{ij},y_{1})G_{\rm c}^{\prime{\rm R}}({\bf r}_{ij},-y_{1}+\omega)
\rho({\bf r}_{ij},y_{1}+\omega)G_{\rm c}^{\prime{\rm R}}({\bf r}_{ij},-y_{1})
\nonumber
\\
& &
\qquad\qquad\qquad\qquad\qquad\quad
-\rho({\bf r}_{ij},-y_{1})G_{\rm c}^{\prime{\rm R}}({\bf r}_{ij},-y_{1}+\omega)
\rho({\bf r}_{ij},y_{1}+\omega)G_{\rm c}^{\prime{\rm R}}({\bf r}_{ij},y_{1})
\biggr].
\label{rates7B2}
\end{eqnarray}

By changing the integration variable from $y_{4}$ to $y_{1}$ in Eq,\ (\ref{rates7A}), 
Im$\Gamma_{\rm I}^{\rm R}(\omega+{\rm i}\delta)$ is transformed to 
\begin{eqnarray}
& &
{\rm Im}\Gamma_{\rm I}^{\rm R}(\omega+{\rm i}\delta)=
 \frac{2\pi V_{\rm dc}^{2}U_{\rm ph}}{\epsilon_{\rm d}^{2}}
\int_{-\infty}^{\infty}{\rm d}y_{1}\,
{\rm th}\left(\frac{y_{1}}{2T}\right)
\rho({\bf r}_{ij},y_{1}+\omega)G_{\rm c}^{\prime{\rm R}}({\bf r}_{ij},-y_{1}+\omega)
\nonumber
\\
& &\qquad\qquad\qquad\qquad
\times
\left[
\rho({\bf r}_{ij},y_{1})G_{\rm c}^{\prime{\rm R}}({\bf r}_{ij},-y_{1})
-\rho({\bf r}_{ij},-y_{1})G_{\rm c}^{\prime{\rm R}}({\bf r}_{ij},y_{1})
\right].
\label{rates7A2}
\end{eqnarray}
It is easy to see that the 
expression of integrand in Eq.\ (\ref{rates7B2}) and Eq.\ (\ref{rates7A2}) are same 
except for the difference of argument $x$ in ${\rm th}(x)$.
Since the Im$\Gamma_{\rm II}(\omega+{\rm i}\delta)$ 
arises twice from the integration along Im$z=\pm\omega_{\nu}$, the  $\omega$-linear term in 
twice of Eq.\ (\ref{rates7B2}) and that in Eq.\ (\ref{rates7A2}) cancels with each other 
in the low temperature region, $T\ll \epsilon_{\rm F}$, 
where $\{{\rm th}[(y_{1}+\omega)/2T]- {\rm th}(y_{1}/2T)\}\approx 2\omega\delta(y_{1})$ 
so that the expression in the bracket in Eq.\ (\ref{rates7A2}) technically vanishes.
Therefore, $2$Im$\Gamma_{\rm II}(\omega+{\rm i}\delta)$+ Im$\Gamma_{\rm I}(\omega+{\rm i}\delta)$ 
is given by twice of the second term in the brace of Eq.\ (\ref{rates7B}).  Namely, 
${\rm Im}{\tilde \Gamma}_{\rm ph}^{\rm R}(\omega+{\rm i}\delta)\equiv 
2{\rm Im}\Gamma_{\rm II}^{\rm R}(\omega+{\rm i}\delta)+
{\rm Im}\Gamma_{\rm I}^{\rm R}(\omega+{\rm i}\delta)$ is given by 
\begin{eqnarray}
& &
{\rm Im}{\tilde \Gamma}_{\rm ph}^{\rm R}(\omega+{\rm i}\delta)=
-\frac{{2}\pi V_{\rm dc}^{2}U_{\rm ph}}{\epsilon_{\rm d}^{2}}
\int_{-\infty}^{\infty}{\rm d}y_{4}\,
{\rm th}\left(\frac{y_{4}-\omega}{2T}\right)
\nonumber
\\
& &
\qquad\qquad\qquad\qquad\qquad
\times
\left[\rho({\bf r}_{ij},y_{4}+\omega)\rho({\bf r}_{ij},-y_{4}+\omega)
G_{\rm c}^{\prime{\rm R}}({\bf r}_{ij},-y_{4})G_{\rm c}^{\prime{\rm R}}({\bf r}_{ij},y_{4})
\right]
\nonumber
\\
& &
\qquad\qquad\quad\quad
=-\frac{{2}\pi V_{\rm dc}^{2}U_{\rm ph}}{\epsilon_{\rm d}^{2}}
\int_{-\infty}^{\infty}{\rm d}y_{4}\,
\left[{\rm th}\left(\frac{y_{4}-\omega}{2T}\right)-{\rm th}\frac{y_{4}}{2T}\right]
\nonumber
\\
& &
\qquad\qquad\qquad\qquad\quad
\times
\left[\rho({\bf r}_{ij},y_{4}+\omega)\rho({\bf r}_{ij},-y_{4}+\omega)
G_{\rm c}^{\prime{\rm R}}({\bf r}_{ij},-y_{4})G_{\rm c}^{\prime{\rm R}}({\bf r}_{ij},y_{4})
\right],
\label{rates7C}
\end{eqnarray}
where, in deriving the second equality, we have used the fact that the function in the bracket is 
an even function in $y_{4}$ so that the term including ${\rm th}(y_{4}/2T)$ vanishes.  

\section{Case of direct overlap of electrons between Tl and Te sites}\label{DirectOverlap}
In the case where the localized state at Tl site extends to the adjacent Te site, the relaxation 
function $\Gamma^{\prime}$, corresponding to ${\tilde \Gamma}_{\rm ph}$ defined by Eq.\ (\ref{rates3}), 
is derived from the Feynman diagram 
shown in Fig.\ \ref{Fig:A3} and its vertical inversion. Its analytic expression is given by

\begin{figure}[h]
\begin{center}
\rotatebox{0}{\includegraphics[angle=0,width=0.5\linewidth]{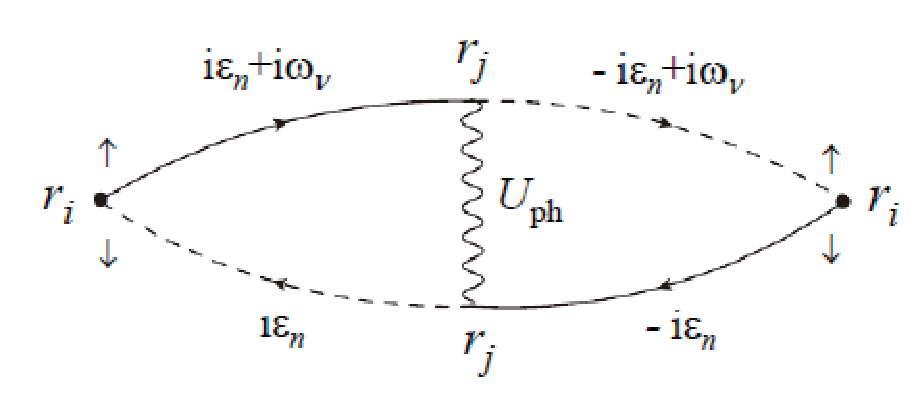}}
\caption{Feynman diagram giving the NMR longitudinal relaxation rates $1/T_{1}T$ at Te (${\bf r}_{i}$) site,  
in the case where 6s electron at  Tl (${\bf r}_{j}$) site extends to the Te (${\bf r}_{i}$) site. }
\label{Fig:A3}
\end{center}
\end{figure}

\begin{eqnarray}
& & 
\Gamma^{\,\prime}({\rm i}\omega_{\nu})
=2U_{\rm ph}T\sum_{\varepsilon_{n}}
G_{\rm c}({\bf r}_{ij},{\rm i}\varepsilon_{n}+{\rm i}\omega_{\nu})G_{\rm c}({\bf r}_{ij},-{\rm i}\varepsilon_{n})
\nonumber
\\
& &\qquad\quad
\qquad\qquad
\times
G_{\rm d}(-{\rm i}\varepsilon_{n}+{\rm i}\omega_{\nu})G_{\rm d}({\rm i}\varepsilon_{n})
\label{rates2A}
\end{eqnarray}
Instead of Eq.\ (\ref{rates3}), we obtain. 
\begin{eqnarray}
\Gamma^{\,\prime}({\rm i}\omega_{\nu})
=2U_{\rm ph}T\sum_{\varepsilon_{n}}
\left(\prod_{\ell=2}^{3}\int_{-\infty}^{\infty}{\rm d}y_{\ell}\right)\,
\frac{\rho({\bf r}_{ij},y_{2})}{{\rm i}\varepsilon_{n}+{\rm i}\omega_{\nu}-y_{2}}\,
\frac{\rho({\bf r}_{ij},y_{3})}{-{\rm i}\varepsilon_{n}-y_{3}}\,
\frac{1}{{\rm i}\varepsilon_{n}-\epsilon_{\rm d}}
\frac{1}{-{\rm i}\varepsilon_{n}+{\rm i}\omega_{\nu}-\epsilon_{\rm d}}
\label{rates8}
\end{eqnarray}

Note that we have not approximated 
$G_{\rm d}(-{\rm i}\varepsilon_{n})G_{\rm d}(-{\rm i}\varepsilon_{n})$ by $1/\epsilon_{\rm d}^{2}$.  
The summation in Eq.\ (\ref{rates8}) with respect to $\varepsilon_{n}$ is performed in a standard way 
by transforming the summation to the integration along the axes Im$z=0$ and Im$z=\pm\omega_{\nu}$ 
on $z$-plane, where one is just above these axes and another is just below in the counter direction. 
The result is 
\begin{eqnarray}
& & 
\Gamma^{\,\prime}({\rm i}\omega_{\nu})
=U_{\rm ph}
\left(\prod_{\ell=2}^{3}\int_{-\infty}^{\infty}{\rm d}y_{\ell}\right)\,
\rho({\bf r}_{ij},y_{2})\rho({\bf r}_{ij},y_{3})
\nonumber
\\
& &
\quad
\times
\Biggl\{
\frac{1}{y_{3}+\epsilon_{\rm d}}
\left[{\rm th}\left(\frac{y_3}{2T}\right)\,
\frac{1}{-y_{2}-y_{3}+{\rm i}\omega_{\nu}}\,
\frac{1}{y_{3}-\epsilon_{\rm d}+{\rm i}\omega_{\nu}}
+
{\rm th}\left(\frac{\epsilon_{\rm d}}{2T}\right)\,
\frac{1}{\epsilon_{\rm d}-y_{2}+{\rm i}\omega_{\nu}}\,
\frac{1}{-2\epsilon_{\rm d}+{\rm i}\omega_{\nu}}
\right]
\nonumber \\
& &
\quad
+\frac{1}{y_{2}+\epsilon_{\rm d}-2{\rm i}\omega_{\nu}}
\left[{\rm th}\left(\frac{y_2}{2T}\right)\,
\frac{1}{-y_{2}-y_{3}+{\rm i}\omega_{\nu}}\,
\frac{1}{y_{2}-\epsilon_{\rm d}-{\rm i}\omega_{\nu}}
+
{\rm th}\left(\frac{\epsilon_{\rm d}}{2T}\right)\,
\frac{1}{\epsilon_{\rm d}-y_{3}-{\rm i}\omega_{\nu}}\,
\frac{1}{-2\epsilon_{\rm d}+{\rm i}\omega_{\nu}}
\right]
\Biggr\}.
\label{rates9}
\end{eqnarray}
After analytic continuation, ${\rm i}\omega_{\nu}\to \omega+{\rm i}\delta$ in Eq.\ (\ref{rates9}), 
and taking an imaginary part, we obtain
\begin{eqnarray}
& & 
{\rm Im}\Gamma^{\,\prime{\rm R}}(\omega+{\rm i}\delta)
=-\pi U_{\rm ph}
\left(\prod_{\ell=2}^{3}\int_{-\infty}^{\infty}{\rm d}y_{\ell}\right)\,
\rho({\bf r}_{ij},y_{2})\rho({\bf r}_{ij},y_{3})\,\frac{1}{y_{3}+\epsilon_{\rm d}}
\nonumber
\\
& &
\qquad\qquad\qquad\qquad\qquad
\times
\Biggl\{
{\rm th}\left(\frac{y_3}{2T}\right)\,
\left[
\frac{\delta(y_{3}-\epsilon_{\rm d}+\omega)}{-y_{2}-y_{3}+\omega}
+\frac{\delta(y_{2}+y_{3}-\omega)}{y_{3}-\epsilon_{\rm d}+\omega}
\right]
\nonumber
\\
& &
\qquad\qquad\qquad\qquad\qquad\qquad
+
{\rm th}\left(\frac{\epsilon_{\rm d}}{2T}\right)\,
\left[
\frac{\delta(-2\epsilon_{\rm d}+\omega)}{\epsilon_{\rm d}-y_{2}+\omega}
+\frac{\delta(\epsilon_{\rm d}-y_{2}+\omega)}{-2\epsilon_{\rm d}+\omega}
\right]
\Biggr\}
\nonumber
\\
& &
\qquad\qquad
+\pi U_{\rm ph}
\left(\prod_{\ell=2}^{3}\int_{-\infty}^{\infty}{\rm d}y_{\ell}\right)\,
\rho({\bf r}_{ij},y_{2})\rho({\bf r}_{ij},y_{3})
\nonumber
\\
& &
\qquad\qquad\qquad\qquad\qquad
\times
\Biggl\{{\rm th}\left(\frac{y_2}{2T}\right)\frac{1}{-y_{2}-\epsilon_{\rm d}+2\omega}
\left[
\frac{\delta(y_{2}+y_{3}-\omega)}{y_{2}-\epsilon_{\rm d}-\omega}
+\frac{\delta(y_{2}-\epsilon_{\rm d}-\omega)}{y_{2}+y_{3}-\omega}
\right]
\nonumber
\\
& &
\qquad\qquad\qquad\qquad\qquad\qquad
-{\rm th}\left(\frac{y_2}{2T}\right)\frac{\delta(y_{2}+\epsilon_{\rm d}-2\omega)}
{(y_{2}+y_{3}-\omega)(y_{2}-\epsilon_{\rm d}-\omega)}
\nonumber
\\
& &
\qquad\qquad\qquad\qquad\qquad\qquad
+
{\rm th}\left(\frac{\epsilon_{\rm d}}{2T}\right)\,
\frac{1}{-2\epsilon_{\rm d}+\omega}
\left[
\frac{\delta(y_{2}+\epsilon_{\rm d}-2\omega)}{\epsilon_{\rm d}-y_{3}-\omega}
+\frac{\delta(y_{3}-\epsilon_{\rm d}+\omega)}{\epsilon_{\rm d}+y_{2}-2\omega}
\right]
\Biggr\}.
\label{rates10}
\end{eqnarray}
Performing the integration with respect to $y_{2}$ or $y_{3}$, Eq.\ (\ref{rates10}) is reduced to 
{
\begin{eqnarray}
& & 
{\rm Im}\Gamma^{\,\prime{\rm R}}(\omega+{\rm i}\delta)
=-\pi U_{\rm ph}
\int_{-\infty}^{\infty}{\rm d}y_{2}\rho({\bf r}_{ij},y_{2})
\nonumber
\\
& &
\qquad\qquad\qquad\quad
\times
\Biggl\{
\left[
{\rm th}\left(\frac{\epsilon_{\rm d}-\omega}{2T}\right)
-{\rm th}\left(\frac{\epsilon_{\rm d}}{2T}\right)
\right]
\frac{\rho({\bf r}_{ij},\epsilon_{\rm d}-\omega)}{(-\epsilon_{\rm d}-y_{2}+2\omega)
(2\epsilon_{\rm d}-\omega)}
\nonumber
\\
& &
\qquad\qquad\qquad\qquad
+\left[{\rm th}\left(\frac{\epsilon_{\rm d}-2\omega}{2T}\right)
-{\rm th}\left(\frac{\epsilon_{\rm d}}{2T}\right)
\right]
\frac{\rho({\bf r}_{ij},-\epsilon_{\rm d}+2\omega)}{(-\epsilon_{\rm d}+y_{2}+\omega)
(2\epsilon_{\rm d}-\omega)}
\nonumber
\\
& &
\qquad\qquad\qquad\qquad
-\left[{\rm th}\left(\frac{y_{2}-\omega}{2T}\right)-{\rm th}\left(\frac{y_{2}}{2T}\right)\right]
\frac{\rho({\bf r}_{ij},-y_{2}+\omega)}
{(-y_{2}-\epsilon_{\rm d}+2\omega)
(-y_{2}+\epsilon_{\rm d}+\omega)}
\nonumber
\\
& &
\qquad\qquad\qquad\qquad
+\left[{\rm th}\left(\frac{\epsilon_{\rm d}}{2T}\right)
-{\rm th}\left(\frac{\epsilon_{\rm d}+\omega}{2T}\right)\right]
\frac{\rho({\bf r}_{ij},\epsilon_{\rm d}+\omega)}
{(y_{2}+\epsilon_{\rm d})(-2\epsilon_{\rm d}+\omega)}
\Biggr\}.
\label{rates11}
\end{eqnarray}
The first and fourth terms give only vanishing contribution because $\rho({\bf r}_{ij},\epsilon_{\rm d})$ 
[Eq.\ (\ref{GF2})] are vanishing in the present case $\epsilon_{\rm d}<-\epsilon_{\rm F}$.  
The second term also gives vanishing contribution because 
$\{{\rm th}[(\epsilon_{\rm d}-2\omega)/2T]-{\rm th}(\epsilon_{\rm d}/2T)\}$ is vanishing 
if the $\epsilon_{\rm d}$ is located well below the bottom of the conduction band 
(in the hole picture).  
}
Therefore, Eq.\ (\ref{rates11}) is finally reduced to    
\begin{eqnarray}
{\rm Im}\Gamma^{\,\prime{\rm R}}(\omega+{\rm i}\delta)
=\pi U_{\rm ph}\int_{-\infty}^{\infty}{\rm d}y_{2}
\left[{\rm th}\left(\frac{y_{2}-\omega}{2T}\right)-{\rm th}\frac{y_{2}}{2T}\right]
\frac{\rho({\bf r}_{ij},y_{2})\rho({\bf r}_{ij},-y_{2}+\omega)}
{(-y_{2}-\epsilon_{\rm d}+2\omega)(-y_{2}+\epsilon_{\rm d}+\omega)}.
\label{rates12}
\end{eqnarray}
Then, up to the linear term in $\omega$, 
${\rm Im}\Gamma^{\,\prime}(\omega+{\rm i}\delta)$ is 
given as
\begin{equation}
{\rm Im}\Gamma^{\,\prime{\rm R}}(\omega+{\rm i}\delta)
\approx -\,\omega\pi U_{\rm ph}\int_{-\infty}^{\infty}{\rm d}y_{2}
\frac{\partial\, {\rm th}\left(\frac{y_{2}}{2T}\right)}{\partial y_{2}}
\frac{\rho({\bf r}_{ij},y_{2})\rho({\bf r}_{ij},-y_{2})}
{(-y_{2}-\epsilon_{\rm d})(-y_{2}+\epsilon_{\rm d})}.
\label{rates13}
\end{equation}
Considering that $\rho({\bf r}_{ij},y_{2})\rho({\bf r}_{ij},-y_{2}+\omega)$ with $\omega\approx 0$ 
is vanishing at $|y_{2}|>\epsilon_{\rm F}$ and $|\epsilon_{\rm d}|\gg\epsilon_{\rm F}$, 
the expression [Eq.\ (\ref{rates13})] is further simplified as 
\begin{eqnarray}
& &
{\rm Im}\Gamma^{\,\prime{\rm R}}(\omega+{\rm i}\delta)
\approx \omega\frac{\pi U_{\rm ph}}{\epsilon_{\rm d}^{2}}\int_{-\infty}^{\infty}{\rm d}y_{2}
\frac{\partial\, {\rm th}\left(\frac{y_{2}}{2T}\right)}{\partial y_{2}}
\rho({\bf r}_{ij},y_{2})\rho({\bf r}_{ij},-y_{2})
\nonumber
\\
& &\qquad\qquad\qquad
\approx
\omega\frac{2\pi U_{\rm ph}}{\epsilon_{\rm d}^{2}}\left[\rho({\bf r}_{ij},0)\right]^{2}.
\label{rates14}
\end{eqnarray}

If  $G_{\rm d}({\rm i}\varepsilon_{n}+{\rm i}\omega_{\nu})
G_{\rm d}(-{\rm i}\varepsilon_{n})$ in Eq.\ (\ref{rates2A}) is approximated by 
$1/\epsilon_{\rm d}^{2}$ as in Eq.\ (\ref{rates3}), the relaxation 
function $\Gamma^{\,\prime\prime}({\rm i}\omega_{\nu})$ is easily calculated as follows: 
\begin{eqnarray}
& &
\Gamma^{\,\prime\prime{\rm R}}({\rm i}\omega_{\nu})
=\frac{2U_{\rm ph}}{\epsilon_{\rm d}^{2}}T\sum_{\varepsilon_{n}}
G_{\rm c}({\bf r}_{ij},{\rm i}\varepsilon_{n}+{\rm i}\omega_{\nu})G_{\rm c}({\bf r}_{ij},-{\rm i}\varepsilon_{n})
\nonumber
\\
& &
\qquad
=\frac{2U_{\rm ph}}{\epsilon_{\rm d}^{2}}T\sum_{\varepsilon_{n}}
\left(\prod_{\ell=2}^{3}\int_{-\infty}^{\infty}{\rm d}y_{\ell}\right)\,
\frac{\rho({\bf r}_{ij},y_{2})}{{\rm i}\varepsilon_{n}+{\rm i}\omega_{\nu}-y_{2}}\,
\frac{\rho({\bf r}_{ij},y_{3})}{-{\rm i}\varepsilon_{n}-y_{3}}\,
\nonumber
\\
& &
\qquad
=\frac{2U_{\rm ph}}{\epsilon_{\rm d}^{2}}
\left(\prod_{\ell=2}^{3}\int_{-\infty}^{\infty}{\rm d}y_{\ell}\right)\,
\frac{1}{2}\left({\rm th}\frac{y_{2}}{2T}+{\rm th}\frac{y_{3}}{2T}\right)
\frac{\rho({\bf r}_{ij},y_{2})\rho({\bf r}_{ij},y_{3})}{-y_{2}-y_{3}+{\rm i}\omega_{\nu}}
\label{rates15}
\end{eqnarray}
After analytic continuation, ${\rm i}\omega_{\nu}\to \omega+{\rm i}\delta$ in Eq.\ (\ref{rates15}), and 
performing an integration with respect to $y_{3}$, 
Im$\Gamma^{\prime\prime{\rm R}}(\omega+{\rm i}\delta)$ is reduced to 
\begin{eqnarray}
{\rm Im}\Gamma^{\,\prime\prime{\rm R}}(\omega+{\rm i}\delta)
=\frac{2\pi U_{\rm ph}}{\epsilon_{\rm d}^{2}}
\int_{-\infty}^{\infty}{\rm d}y_{2}\,
\frac{1}{2}\left[{\rm th}\frac{y_{2}}{2T}-{\rm th}\left(\frac{y_{2}-\omega}{2T}\right)\right]
\rho({\bf r}_{ij},y_{2})\rho({\bf r}_{ij},-y_{2}+\omega).
\label{rates16}
\end{eqnarray}
Then, up to the linear order in $\omega$, 
${\rm Im}\Gamma^{\,\prime\prime}(\omega+{\rm i}\delta)$ is 
given as
\begin{eqnarray}
& &
{\rm Im}\Gamma^{\,\prime\prime{\rm R}}(\omega+{\rm i}\delta)
\approx \omega\frac{\pi U_{\rm ph}}{\epsilon_{\rm d}^{2}}\int_{-\infty}^{\infty}{\rm d}y_{2}
\frac{\partial\, {\rm th}\left(\frac{y_{2}}{2T}\right)}{\partial y_{2}}
\rho({\bf r}_{ij},y_{2})\rho({\bf r}_{ij},-y_{2})
\nonumber
\\
& &
\qquad\qquad\qquad
\approx
\omega\frac{2\pi U_{\rm ph}}{\epsilon_{\rm d}^{2}}
\left[\rho({\bf r}_{ij},0)\right]^{2},
\label{rates17}
\end{eqnarray}
which is the same as the expression [Eq.\ (\ref{rates14})]. This justifies the approximation 
$G_{\rm d}({\rm i}\varepsilon_{n}+{\rm i}\omega_{\nu})
G_{\rm d}(-{\rm i}\varepsilon_{n})\approx 1/\epsilon_{\rm d}^{2}$ in Eq.\ (\ref{rates2A}), which in turn 
justifies the same approximation adopted in Eq.\ (\ref{rates3}).  
A physical basis of this justification is that the 
${\rm Im}\Gamma^{{\prime}{\rm R}}(\omega+{\rm i}\delta)$ arises only from the low energy processes 
associated with conduction electrons described by 
$G_{\rm c}({\bf r}_{ij},{\rm i}\varepsilon_{n}+{\rm i}\omega_{\nu})$ and 
$G_{\rm c}({\bf r}_{ij},-{\rm i}\varepsilon_{n})$ in Eq.\ (\ref{rates2A}), so that the same approximation 
is expected to remain valid also in the calculation of  
${\rm Im}{\tilde \Gamma}_{\rm ph}^{\rm R}(\omega+{\rm i}\delta)$ performed in Appendix \ref{Gamma}. 

In the limit $k_{\rm F}|{\bf r}_{ij}|\ll 1$, with the use of asymptotic form of $\rho({\bf r},y)$ 
[Eq.\ (\ref{GF3})], Eq.\ (\ref{rates17}) is estimated as 
\begin{equation}
{\rm Im}\Gamma^{\,\prime\prime{\rm R}}(\omega+{\rm i}\delta)
\approx
\omega\frac{2\pi N_{\rm F}^{2}}{\epsilon_{\rm d}^{2}}e^{-(|{\bf r}_{ij}|/\ell)}U_{\rm ph}\,
+{\cal O}(\omega^{2}).
\label{rates18}
\end{equation}
Then, the NMR relaxation rate $1/T_{1}T$ is given by 
\begin{equation}
\frac{1}{T_{1}T}=A^{2}\frac{2\pi N_{\rm F}^{2}}{\epsilon_{\rm d}^{2}}e^{-(|{\bf r}_{ij}|/\ell)}U_{\rm ph}.
\label{rates19}
\end{equation}

\section{Real-Part of Retarded Green Function of Conduction Electrons}\label{RealPart}
In this Appendix, we derive an analytic form of $G_{\rm c}^{\prime{\rm R}}({\bf r},\varepsilon)$ 
[Eq.\ (\ref{GF1B})] in the limit $k_{\rm F}r\ll 1$, where $G_{\rm c}^{\prime{\rm R}}({\bf r},\varepsilon)$ is 
approximated by 
\begin{equation} 
G_{\rm c}^{\prime{\rm R}}({\bf r},\varepsilon)\approx
-\frac{mk_{\rm F}}{2\pi^{2}}\,e^{-(r/2\ell)}\frac{1}{\sqrt{\epsilon_{\rm F}}}
\int_{-\epsilon_{\rm F}}^{\epsilon_{\rm c}}{\rm d}y\frac{\sqrt{y+\epsilon_{\rm F}}}{y-\varepsilon}.
\label{C1}
\end{equation}
Integration with respect to $y$ is performed by elementary integral leading to the following results. 
In the case $\varepsilon+\epsilon_{\rm F}>0$, 
\begin{equation}
\int_{-\epsilon_{\rm F}}^{\epsilon_{\rm c}}{\rm d}y\frac{\sqrt{y+\epsilon_{\rm F}}}{y-\varepsilon}=
2\sqrt{\epsilon_{\rm c}+\epsilon_{\rm F}}
+\sqrt{\varepsilon+\epsilon_{\rm F}}\,
\log\left|\frac{\sqrt{\epsilon_{\rm c}+\epsilon_{\rm F}}-\sqrt{\varepsilon+\epsilon_{\rm F}}}
{\sqrt{\epsilon_{\rm c}+\epsilon_{\rm F}}+\sqrt{\varepsilon+\epsilon_{\rm F}}}
\right|,
\label{C2}
\end{equation}
while in the case  $\varepsilon+\epsilon_{\rm F}<0$, 
\begin{equation}
\int_{-\epsilon_{\rm F}}^{\epsilon_{\rm c}}{\rm d}y\frac{\sqrt{y+\epsilon_{\rm F}}}{y-\varepsilon}=
2\sqrt{\epsilon_{\rm c}+\epsilon_{\rm F}}
-2\sqrt{-\varepsilon-\epsilon_{\rm F}}\,\tan^{-1}
\frac{\sqrt{\epsilon_{\rm c}+\epsilon_{\rm F}}}{\sqrt{-\varepsilon-\epsilon_{\rm F}}}.
\label{C3}
\end{equation}

\section{Calculation of $J(k_{\rm F}r)$ in the limit $k_{\rm F}r\gg 1$}\label{JkFr}
In this Appendix, we derive an asymptotic form of $J(k_{\rm F}r)$, Eq.\ (\ref{GF1D}), 
in the limit $k_{\rm F}r\gg 1$.  The integration in Eq.\ (\ref{GF1D}) 
with respect to $y$, which is denoted by $K$, is transformed, 
by changing the integration variable from $y$ to $u\equiv\sqrt{(y/\epsilon_{\rm F})+1}$ 
and defining $\Lambda\equiv\sqrt{(\epsilon_{\rm c}/\epsilon_{\rm F})+1}$ , 
as follows: 
\begin{eqnarray}
& &
K=\int_{0}^{\Lambda}{\rm d}u\frac{2u}{(u+1)(u-1)}\sin\left[(k_{\rm F}r)u\right]
\nonumber
\\
& &
\quad
={\rm Im}\left[\int_{0}^{\Lambda}{\rm d}u\frac{2u}{(u+1)(u-1)}e^{{\rm i}(k_{\rm F}r)u}\right], 
\label{D1}
\end{eqnarray}
where the integration with respect to $u$ is the principal integration for avoiding the singularity around $u=1$.  

\begin{figure}[h]
\begin{center}
\rotatebox{0}{\includegraphics[angle=0,width=0.5\linewidth]{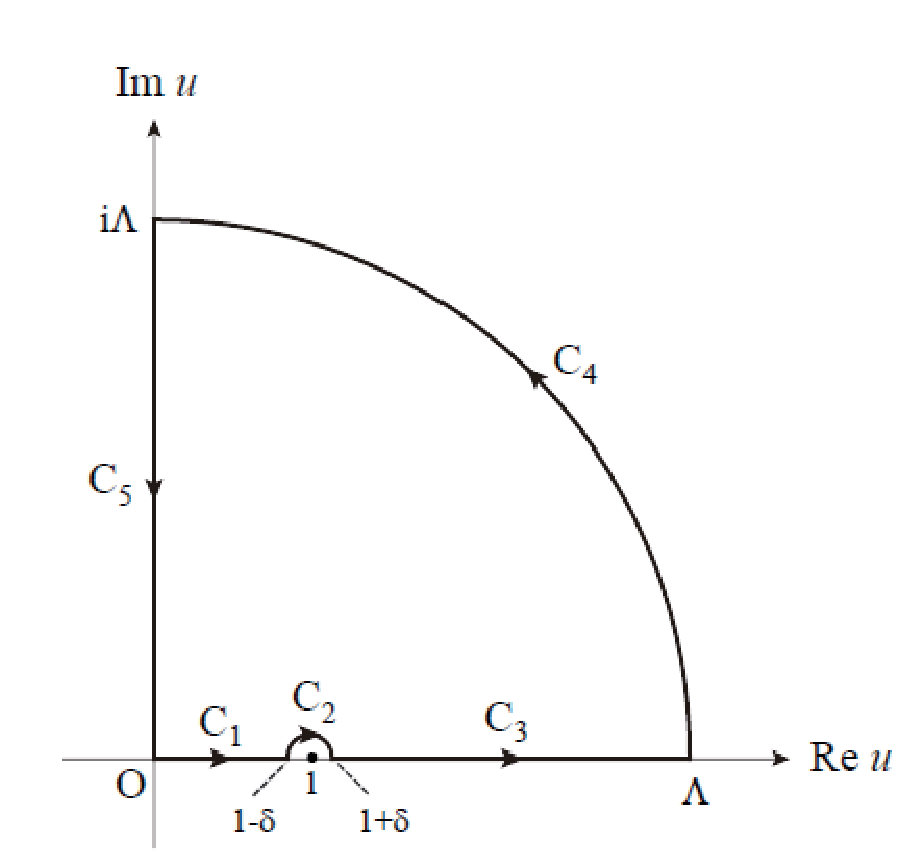}}
\caption{Path of contour integration in Eq.\ (\ref{D1}) in the complex-$u$ plane.}
\label{Fig:A4}
\end{center}
\end{figure}

Let us define $K_{i}$ ($i=1\sim5$)  by integration with respect complex $u$ along the path $C_{i}$ shown 
in Fig.\ \ref{Fig:A4} as
\begin{equation}
K_{i}\equiv\int_{C_{i}}{\rm d}u\frac{2u}{(u+1)(u-1)}e^{{\rm i}(k_{\rm F}r)u}. 
\label{D2}
\end{equation}
An infinitesimally small positive number $\delta$ in Fig.\ \ref{Fig:A4} will be tended 
to zero after calculations.  
$\lim_{\delta\to 0}[K_{1}(\delta)+K_{3}(\delta)]$ is the same as the principal integration in Eq.\ (\ref{D1}). 
The integration along $C_{2}$, a semicircle with the radius $\delta$,  is 
performed in the limit $\delta \to 0$ as 
\begin{eqnarray}
& &
K_{2}=\int_{\pi}^{0}{\rm d}(\delta e^{{\rm i}\varphi})\,
\frac{2(1+\delta e^{{\rm i}\varphi})}{(2+\delta e^{{\rm i}\varphi})\,\delta e^{{\rm i}\varphi}}
e^{{\rm i}(k_{\rm F}r)(1+\delta e^{{\rm i}\varphi})}
\nonumber
\\
& &
\quad\,\,\,
\approx
{\rm i}\int_{\pi}^{0}{\rm d}\varphi e^{{\rm i}(k_{\rm F}r)}
=-{\rm i}\,\pi\cos\,(k_{\rm F}r)+\pi\sin\,(k_{\rm F}r).
\label{D3}
\end{eqnarray} 
It is easy to see that $K_{5}(\Lambda)$ is real and finite number.  
The integration along $C_{{4}}$, a semicircle with the radius $\Lambda$,  is performed as 
\begin{eqnarray}
& &
K_{4}(\Lambda)=\int_{0}^{\pi/2}{\rm d}(\Lambda e^{{\rm i}\theta})\,
\frac{2\Lambda e^{{\rm i}\theta}}{\Lambda^{2}e^{2{\rm i}\theta}-1}
e^{[{\rm i}(k_{\rm F}r)\Lambda e^{{\rm i}\theta}]}
\nonumber
\\
& &
\qquad\,\,\,\,
={\rm i}\int_{0}^{\pi/2}{\rm d}\theta\,
\frac{2\Lambda^{2} e^{2{\rm i}\theta}}{\Lambda^{2}e^{2{\rm i}\theta}-1}
\,e^{{\rm i}(k_{\rm F}r)\Lambda \cos\theta}
\,e^{-(k_{\rm F}r)\Lambda \sin\theta}.
\label{D3A}
\end{eqnarray} 
It is shown by a standard way of calculus that $K_{4}(\Lambda)$ vanishes in proportion to $1/(k_{\rm F}r)$ 
in the limit $k_{\rm F}r\gg 1$.   

Therefore, Eq.\ (\ref{D1}) is transformed  in the limit $k_{\rm F}r\gg 1$ as follows:
\begin{eqnarray}
& &
K=\lim_{\delta\to 0}{\rm Im} [K_{1}(\delta)+K_{{3}}(\delta)]
\nonumber
\\
& &
\quad\,
=\lim_{\delta\to 0}
{\rm Im}\Biggl\{\left[K_{1}(\delta)+K_{3}(\delta)+K_{2}(\delta)+K_{4}+K_{5} \right]
\nonumber
\\
& &
\qquad\qquad\qquad\qquad
-\left[K_{2}(\delta)+K_{4}+K_{5}\right]
\Biggr\}
\label{D4}
\\
& &
\quad\,
=
-\lim_{\delta\to 0}{\rm Im}\left[K_{2}(\delta)+K_{4}+K_{5}\right]
\label{D5}
\\
& &
\quad\,
\approx
-\lim_{\delta\to 0}{\rm Im}K_{2}(\delta)=\pi\cos\,(k_{\rm F}r).
\label{D6}
\end{eqnarray} 
In deriving Eq.\ (\ref{D5}) from Eq.\ (\ref{D4}), we have used the fact that 
the contour integration in the complex-$u$ plane along the path shown 
in Fig.\ \ref{Fig:A4} vanishes because the integrand is an analytic function 
in the domain encircled by the contour.  
In deriving Eq.\ (\ref{D6}) from Eq.\ (\ref{D5}), we have used Eq.\ (\ref{D3}). 

As a result, $J(k_{\rm F}r)$ [Eq.\ (\ref{GF1D})] in the limit $k_{\rm F}r\gg 1$ is 
given by 
\begin{equation}
J(k_{\rm F}r)\approx \frac{1}{k_{\rm F}r}\pi\cos\,(k_{\rm F}r).
\label{D7}
\end{equation}

\section{Poorman's Scaling Analysis for $U_{\rm ph}$ and $U_{\rm dc}$}\label{Scaling}
In this Appendix, we perform the poorman's scaling analysis for the pair-hopping interaction 
$U_{\rm ph}$ and the inter-orbital interaction $U_{\rm dc}$ to investigate renormalization effect
on these interaction. As discussed in Appendix\ref{equivalence}, in the mapped world, $U_{\rm ph}$ and 
$U_{\rm dc}$ correspond to $J_{\perp}/2$ and $J_{z}/4$ in the anisotropic s-d model. 
The evolution equations for these dimensionless coupling constants, $y_{\perp}\equiv J_{\perp}N_{\rm F}$ 
and  $y_{z}\equiv J_{z}N_{\rm F}$ are given as follows:~\cite{PoormanScaling}
\begin{eqnarray}
& &
\frac{{\rm d}y_{\perp}}{{\rm d}x}=-y_{\perp}y_{z},
\label{E1}
\\
& &
\frac{{\rm d}y_{z}}{{\rm d}x}=-y_{\perp}^{2},
\label{E2}
\end{eqnarray}
where $x\equiv \log(E_{\rm c}/E_{\rm c}^{0})$ with $E_{\rm c}$ and $E_{\rm c}^{0}$ being 
the renormalized and bare bandwidths, respectively. It is well known that 
$y_{\perp}^{2}-y_{z}^{2}={\rm const.}\equiv C$. Substituting $y_{\perp}^{2}=y_{z}^{2}+C$ into 
Eq.\ (\ref{E2}), the evolution equation of $y_{z}$ [Eq.\ (\ref{E2})] is reduced to
\begin{equation}
\frac{{\rm d}y_{z}}{{\rm d}x}=-\left(y_{z}^{2}+C\right).
\label{E3}
\end{equation}
The solution of this differential equation is easily solved: In the case $C\equiv a^{2}>0$, 
\begin{equation}
y_{z}(x)=\frac{y_{z}^{0}-a \tan(ax)}{\displaystyle 1+\frac{y_{z}^{0}}{a}\tan(ax)},
\label{E4}
\end{equation}
where $y_{z}^{0}$ is the initial value of $y_{z}$ at $x=0$. Similarly, in the case $C\equiv -b^{2}<0$, 
the solution is given as 
\begin{equation}
y_{z}(x)=\frac{y_{z}^{0}+b \tanh(bx)}{\displaystyle 1+\frac{y_{z}^{0}}{b}\tanh(bx)}.
\label{E5}
\end{equation}
In the high temperature region, $T\gsim T_{\rm K}$, where $|x|\ll 1$, both $y_{z}(x)$ [Eq.\ (\ref{E4})] and 
$y_{z}(x)$ [Eq.\ (\ref{E5})] are expressed as 
\begin{equation}
y_{z}(x)\approx y_{z}^{0}-(y_{\perp}^{0})^{2}x+\cdots.
\label{E6}
\end{equation}
With the use of this approximate expression, that for $y_{\perp}$ is easily obtained in the following form 
\begin{equation}
y_{\perp}(x)\approx y_{\perp}^{0}-{y_{\perp}^{0}}y_{z}^{0}x+\cdots.
\label{E7}
\end{equation}

Therefore, in the high temperature region $T\gsim T_{\rm K}$, temperature dependence of 
$U_{\rm ph}=(J_{\perp}/2)$ and 
$U_{\rm dc}=(J_{z}/4)$ are given as follows:
\begin{eqnarray}
& &
U_{\rm ph}(T)=\frac{1}{2N_{\rm F}}\,y_{\perp}\left(\log\frac{T}{E_{\rm c}^{0}}\right)
\approx \frac{1}{2}\left[2U_{\rm ph}^{0}
-8N_{\rm F}U_{\rm ph}U_{\rm dc}^{0}\,\log\frac{T}{E_{\rm c}^{0}}\right],
\label{E8}
\\
& &
U_{\rm dc}(T)=\frac{1}{4N_{\rm F}}\,y_{z}\left(\log\frac{T}{E_{\rm c}^{0}}\right)
\approx \frac{1}{4}\left[4U_{\rm dc}^{0}-N_{\rm F}(2U_{\rm ph}^{0})^{2}\,\log\frac{T}{E_{\rm c}^{0}}\right],
\label{E9}
\end{eqnarray}
where $U_{\rm ph}^{0}$ and $U_{\rm dc}^{0}$ are bare couplings. 
Namely, both $U_{\rm ph}(T)$ and $U_{\rm dc}(T)$ exhibit logarithmic increase toward 
$T=T_{\rm K}$ at which $y_{\perp}(x)$ and 
$y_{z}(x)$ diverges at the level of approximation of the poorman's scaling.~\cite{PoormanScaling} 

 {
On the other hand, both $y_{\perp}(E)$ and $y_{z}(E)$ diverge toward $E=T_{\rm K}$ as 
\begin{equation}
y_{\perp}(E)=\frac{y_{\perp}(0)}{\displaystyle 1+y_{\perp}(0)\log\frac{E}{E_{\rm c}^{0}}}
=\frac{1}{\displaystyle \log\frac{E}{T_{\rm K}}}\approx y_{z}(E),
\label{E10}
\end{equation}
where $T_{\rm K}$ is 
given by the solution  for the case $C=0$ as $T_{\rm K}=E_{\rm c}^{0}e^{-y_{\perp}(0)}$ or 
$[1+y_{\perp}(0)\log(T_{\rm K}/E_{\rm c}^{0})]=0$.  Of course, this divergence at $E=T_{\rm K}$ 
is an artifact due to the one-loop order approximation, but true divergence occurs in the limit 
$E\ll T_{\rm K}$. Namely, the expression [Eq.(\ref{E10})] is not valid very near at $E=T_{\rm K}$ 
while it gives growing tendency of $y_{\perp}(E)$ and $y_{z}(E)$ toward $E=T_{\rm K}$
}


\end{document}